\documentclass [12pt,a4paper]{article}
\usepackage{bbm}
\usepackage{amsfonts}
\usepackage{mathrsfs}
\usepackage{booktabs}
\usepackage {amssymb}
\usepackage {amsmath}
\usepackage{amsthm}
\usepackage{latexsym}
\usepackage{float}
\usepackage{enumerate}

\def\dse#1{\vskip 0.6cm\noindent
        {\large\bf #1}
        \vskip 0.4cm}

\def\dse#1{\vskip 0.6cm\noindent
        {\large\bf #1}
        \vskip 0.4cm}
\usepackage{lastpage}
\usepackage{epsfig}

\begin{document}
\begin{center}
\textbf{Several new classes of optimal $p$-ary cyclic codes}\footnote {~Corresponding author.

~~~Email addresses: mengenfang@stu.ahau.edu.cn(M.~Fang), lilanqiang716@126.com(L. Li), tianfuyin0825@163.com(F. Tian), liuli-1128@163.com(L. Liu).}
\end{center}

\begin{center}
{ {Mengen Fang~\textsuperscript{a}, \  Lanqiang Li~\textsuperscript{a,\textasteriskcentered}, \  Fuyin Tian~\textsuperscript{a}, \  Li Liu~\textsuperscript{b}} }
\end{center}

\begin{center}
\textit{\textsuperscript{a} \footnotesize  School of Information and Artificial Intelligence, Anhui Agricutural University, Hefei 230036, Anhui, P.R.China }\\
\textit{\textsuperscript{b} \footnotesize  School of Mathematics, Hefei University of Technology, Hefei 230009, Anhui, P.R.China }
\end{center}

\noindent\textbf{Abstract} Cyclic codes, as a crucial subclass of linear codes, exhibit broad applications in communication systems, data storage systems, and consumer electronics, primarily attributed to their well-structured algebraic properties. Let $p$ denote an odd prime with $p\geq5$, and let $m$ be a positive integer. The primary objective of this paper is to construct three novel classes of optimal $p$-ary cyclic codes, denoted as ${\mathcal{C}_p}(0,s,t)$, which possess the parameters $[{p^m} - 1,{p^m} - 2m - 2,4]$. Here, $s$ is defined as $s = \frac{{{p^m}+1}}{{2}}$, and $t$ satisfies the condition $2 \le t \le {p^m} - 2$. Notably, one of the constructed classes includes certain known optimal quinary cyclic codes as special cases. Furthermore, for the specific case when $p=5$, this paper additionally presents four new classes of optimal cyclic codes ${\mathcal{C}_5}(0,s,t)$.\\



\dse{1~~Introduction}
Let $\mathbb{F}_p$ denote the finite field of order $p$, where $p>2$ is a prime. Set $n=p^m-1$, where $\gcd(n,p)=1$. A linear code $\mathcal{C}$ with parameters $[n, k, d]$ over $\mathbb{F}_p$ is defined as a $k$-dimensional subspace of $\mathbb{F}_p^n$ with minimum Hamming distance $d$. Cyclic codes form a subclass of linear codes. A linear code $\mathcal{C}$ is called cyclic if, for every codeword $\mathbf{c} = (c_0, c_1, \cdots, c_{n-1}) \in \mathcal{C}$, every cyclic shift of $\mathbf{c}$ is also contained in $\mathcal{C}$. This combinatorial property gives rise to an elegant algebraic characterization: Codewords $(c_0, c_1, \cdots, c_{n-1})$ in $\mathcal{C}$ correspond bijectively to polynomials of the form $\sum_{i=0}^{n-1} c_i x^i$, which form an ideal in the quotient ring $\mathbb{F}_p[x]/(x^n - 1)$. The generator polynomial of $\mathcal{C}$ is the monic polynomial $g(x)$ of minimal degree that divides $(x^n - 1)$ and belongs to $\mathcal{C}$. Let $\beta \in \mathbb{F}_{p^m}^*$ be a primitive $n$-th root of unity. For $0\leq j \leq n-1$, denote by $m_j(x)$ the minimal polynomial of $\beta^j$ over $\mathbb{F}_p$. We say that the code $\mathcal{C}$ has $l$ zeros $\beta^{j_1},\beta^{j_2},\cdots,\beta^{j_l} $ if the generator polynomial $g(x)$ factors as the product of $l$ pairwise distinct minimal polynomials $m_{j_1}(x),m_{j_2}(x),\cdots,m_{j_l}(x)$ in $\mathbb{F}_p[x]$, where $l$ is an integer satisfying $0\leq l \leq n-1$. Specifically, we denote by $\mathcal{C}_p(j_1,j_2,\cdots,j_l)$ the $p$-ary cyclic code with generator polynomial $m_{j_1}(x)m_{j_2}(x) \cdots m_{j_l}(x)$.

A significant number of optimal cyclic codes have been developed over the past few years. The construction of optimal cyclic codes, particularly ternary codes, has been an active research area. Seminal work by Carlet, Ding and Yuan [2] in 2005 introduced several distance-optimal ternary cyclic codes $\mathcal{C}_3(1,t)$ with two zeros by utilizing perfect nonlinear monomials, which laid a foundation for subsequent research on optimal cyclic codes. In 2013, Ding and Helleseth [3] employed almost perfect nonlinear monomials and other tools over $\mathbb{F}_{3^m}$ to unveil various classes of optimal ternary cyclic codes achieving minimum distance $d=4$. Their work concluded by posing nine open problems, which have since motivated extensive research [5,8,13,15] aimed at solving these problems and exploring other types of optimal ternary cyclic codes.

However, optimal $p$-ary cyclic codes ($p\ge 5$) also constitute a critical topic in coding theory. Compared with binary and ternary cyclic codes, constructing optimal cyclic codes over higher-order fields presents greater challenges. In 2016, Xu, Cao and Xu [6] presented two classes of optimal $p$-ary cyclic codes $\mathcal{C}_p(0,1,t)$ by utilizing perfect nonlinear monomials and the inverse function over $\mathbb{F}_{p^m}$. They also constructed several classes of optimal quinary cyclic codes $\mathcal{C}_5(0,1,t)$ by applying almost perfect nonlinear monomials and other monomials over $\mathbb{F}_{5^m}$. Later, Liu and Cao [11] not only studied the construction of the quinary cyclic code $\mathcal{C}_5(0,1,t)$, but also presented two specific constructions of the optimal quinary cyclic code $\mathcal{C}_5(1,t,\frac{{5^m}-1}{2})$. In 2023, by utilizing solutions to specific equations over finite fields, Wu, Liu and Zhang [16] constructed four classes of optimal $p$-ary cyclic codes $\mathcal{C}_p(1,t,\frac{{p^m}-1}{2})$. To further extend these results, they employed irreducible factors over the finite field $\mathbb{F}_{5^m}$ developed two additional classes of optimal quinary cyclic codes $\mathcal{C}_5(1,t,\frac{{5^m}-1}{2})$. During the same year, Wu, Liu and Li [18] established sufficient conditions for 7-ary cyclic codes $\mathcal{C}_7(0,1,t)$ to be optimal  and presented three classes of such optimal codes. In 2025, Zha, Hu and Wu [21] constructed six classes of optimal $p$-ary cyclic codes $\mathcal{C}_p(1,t,\frac{{p^m}-1}{2})$ and one class of optimal quinary cyclic codes $\mathcal{C}_5(1,t,\frac{{5^m}-1}{2})$. That same year, Fan and Zeng [20] established necessary and sufficient conditions for a quinary cyclic code $\mathcal{C}_5(0,\frac{{5^m}+1}{2},t)$ to be optimal. Using these conditions, they further presented several classes of quinary cyclic codes $\mathcal{C}_5(0,\frac{{5^m}+1}{2},t)$. Most recently, Liu, Cao and Zha [22] constructed eight classes of optimal quinary cyclic codes $\mathcal{C}_5(1,t,\frac{5^m-1}{2})$ through the analysis of solutions to specific equations over $\mathbb{F}_{5^m}$. These collective advancements have significantly inspired our current work.

In this work, we primarily investigate three classes of optimal $p$-ary cyclic codes $\mathcal{C}_p(0,s,t)$ for general odd primes $p \ge 5$, where $s=\frac{{p^m}+1}{2}$. The investigation is conducted by employing case analysis based on the quadratic character of finite fields. Furthermore, we present explicit constructions of specific optimal quinary cyclic codes $\mathcal{C}_5(0,\frac{{5^m}+1}{2},t)$ through examining solutions to specific equations over the finite field $\mathbb{F}_{5^m}$.

The remainder of this work is organized as follows. Section 2 presents fundamental lemmas that are indispensable for the subsequent analysis. Section 3 proposes three new constructions of optimal $p$-ary cyclic codes $\mathcal{C}_p(0,\frac{{p^m}+1}{2},t)$ over $\mathbb{F}_p$. Further, Section 4 presents four classes of optimal quinary cyclic codes $\mathcal{C}_5(0,\frac{{5^m}+1}{2},t)$. Finally, Section 5 offers closing comments and summarizes the main contributions.

\dse{2~~Preliminaries}
  Let $p$ be an odd prime and $m$ be a positive integer. For any integer $u$, the $p$-cyclotomic coset of $u$ modulo ${p^m} - 1$ is given by
  $${C_u} = \{ u,up, \cdots u{p^{{l_u} - 1}}\bmod ({p^m} - 1)\},$$
   where ${l_u}$ denotes the smallest positive integer such that $u \cdot {p^{{l_u}}} \equiv u\bmod ({p^m} - 1)$. The cardinality of ${C_u}$ is $|{C_u}| = {l_u}$. Clearly, $1 \le {l_u} \le m$ and ${l_u}\mid m$ (i.e., $l_u$ divides $m$).

Given an arbitrary codeword $\mathbf{x}=(x_0,x_1,\cdots ,x_{n-1})\in \mathbb{F}_{p^m}^n$, the Hamming weight $\omega_H(x)$ is defined as the cardinality of the set $\{i|x_i \ne0,~0\le i\le n-1\}$.

For any two codewords $\mathbf{x}=(x_0,x_1,\cdots ,x_{n-1})$ and $\mathbf{y}=(y_0,y_1,\cdots ,y_{n-1})$ in $\mathbb{F}_{p^m}^n$, the Hamming distance $d_H(\mathbf{x},\mathbf{y})$ is defined as the number of positions where the corresponding components of $\mathbf{x}$ and $\mathbf{y}$ differ. Formally, it is expressed as:
$$d_H(\mathbf{x},\mathbf{y})=\sum_{i=0}^{n-1} \sigma (x_i\ne y_i),$$
where $\sigma(\cdot)$ denotes the indicator function: $\sigma(\text{true})=1$ (i.e., $\sigma(x_i\ne y_i)=1$ if $x_i\ne y_i$) and $\sigma(\text{false})=0$ (i.e., $\sigma(x_i\ne y_i)=0$ if $x_i=y_i$).
   
Throughout this paper, we denote the quadratic character over $\mathbb{F}_{p^m}$ by $\mu$, which is defined as follows: $\mu(0)=0$ (for the zero element of $\mathbb{F}_{p^m}$); $\mu(v)=1$ if $v$ is a square element in $\mathbb{F}_{p^m}^*$ (i.e.,$v=w^2$ for some $w\in\mathbb{F}_{p^m}^*$); $\mu(v)=-1$ if $v$ is a nonsquare element in $\mathbb{F}_{p^m}^*$ (i.e., there exist no $w\in\mathbb{F}_{p^m}^*$ such that $v=w^2$).

The subsequent two lemmas play a vital role in calculating the cardinality of the cyclotomic coset $C_t$. Consequently, they are essential for determining the dimension of $\mathcal{C}_p(0,s,t)$.\\

\noindent\textbf{Lemma 2.1.} ([5]) Let $h$, $l$ and $r$ be positive integers. Then
$\gcd ({p^h} - 1,{p^r} - 1) = {p^{\gcd (h,r)}} - 1$, and 
\begin{eqnarray*}
\gcd ({p^h} - 1,{p^l} + 1)=
\left\{ {{\begin{array}{ll}
 2, & {\textrm{if }\frac{h}{\gcd (h,l)}\textrm{ is odd}},\\
 {p^{\gcd (h,l)}} + 1, &{\textrm{if }\frac{h}{\gcd (h,l)}\textrm{ is even}}. \\
\end{array} }} \right .
\end{eqnarray*}

\noindent\textbf{Lemma 2.2.} ([6, Lemmas 1, 2]) Let $n = {p^m} - 1$. For any integer $t$ satisfying $1 \le t \le {p^m}-2$, the cardinality $ |{C_t}| $ equals $m$ if either of the following conditions holds:

1) $0< \gcd (t,{p^m} - 1)< p$.

2) $\gcd (t,{p^m} - 1)\gcd ({p^i} - 1,{p^m} - 1) \not\equiv 0~(\bmod {p^m} - 1)$ for all integers $i$ with $0< i < m$. \\

Moreover, the following results will be essential throughout the subsequent proofs.\\ 

\noindent\textbf{Lemma 2.3.} ([1, Theorem 3.46]) Let $f(x)\in{\mathbb{F}_q}[x]$ be an irreducible polynomial of degree $a$. For any integer $b$, $f(x)$ factors into $\gcd (a,b)$ irreducible polynomials in $\mathbb{F}_{{q^b}}[x]$, each of degree $\frac{a}{\gcd (a,b)}$.\\

\noindent\textbf{Lemma 2.4.} ([1, Corollary 3.47]) An irreducible polynomial over $\mathbb{F}_q$ of exact degree $r$ remains irreducible over $\mathbb{F}_{q^l}$ if and only if $r$ and $l$ are coprime.\\

\noindent\textbf{Lemma 2.5.} ([21, Lemma 5]) For a polynomial $f(y)=\alpha y^2+\beta y+\gamma\in\mathbb{F}_p[y]$ with $\alpha \neq0$ or $\beta \neq0$, all its roots lie in $\mathbb{F}_{p^2}$.\\

\noindent\textbf{Lemma 2.6.} ([21, Lemma 6]) Let $q$ is an odd prime power of the form $q=p^i$, where $p$ is a prime and $i$ is a positive integer. For any $\alpha\in \mathbb{F}_q^*$, the following holds:

1) If $m\equiv 0~(\bmod~2)$, then $\alpha^{\frac{q^m-1}{2}}=1$.

2) If $m\equiv 0~(\bmod~4)$, then $\alpha^{\frac{q^m-1}{4}}=1$.\\

\dse{3~~ Optimal $p$-ary cyclic codes $\mathcal{C}_p(0,s,t)$ }

We investigate three new classes of optimal $p$-ary cyclic codes $\mathcal{C}_p(0,s,t)$ with parameters $[p^m- 1,p^m - 2m - 2,4]$, where $s=\frac{p^m+1}{2}.$\\ 

 \noindent\textbf{Lemma 3.1.} Let $s = \frac{p^m + 1}{2}$, where $p\ge 5$ is an odd prime. Then $|{C_s}| = m$. \\

\noindent\textbf{Proof.} Since $\gcd(p^m+1,p^m-1)=2$, it follows that $\gcd(s,p^m-1)=\gcd(\frac{p^m+1}{2},p^m-1) \le 2< p$. By Lemma 2.2, we have $|{C_s}|=m$.\qed\\

To investigate the optimality of the subcode $\mathcal{C}_p(0,s,t)$ of $\mathcal{C}_p(0,s)$, we first determine the minimum Hamming distance $d$ of $\mathcal{C}_p(0,s)$.\\

\noindent\textbf{Lemma 3.2.} Let $p \ge 5$ be an odd prime and $s=\frac{p^m+ 1}{2}$. Then the $p$-ary cyclic code $\mathcal{C}_p(0,s)$ is a $[p^m - 1,{p^m} - m - 2,d]$ code, where $d\ge3$ if $p^m\equiv1~(\textrm{mod }4)$, and $d=2$ if $p^m\equiv3~(\textrm{mod }4)$.\\

 \noindent\textbf{Proof.} Clearly, $\mathcal{C}_p(0,s)$ contains no codeword of Hamming weight 1. Suppose, for contradiction, that $\mathcal{C}_p(0,s)$ has a codeword of Hamming weight 2. Then there exist scalars 
  $c_1,c_2$ in $\mathbb{F}_p^*$ and distinct elements $x_1,x_2$ in $\mathbb{F}_{p^m}^*$ such that
\begin{align}\label{eq:3.1}
\left\{ \begin{array}{l}
c_1 + c_2 = 0,\\
c_1x_1^s + c_2x_2^s = 0.
\end{array} \right.
\end{align}
It is immediate that $c_1 = -c_2$. Since $s =\frac{p^m + 1}{2}= \frac{p^m - 1}{2} + 1$, we note that $x^s =  \pm x$ for any $x\in\mathbb{F}_{p^m}^*$ (this follows from $x^{p^m - 1}=1$, so $x^{\frac{p^m - 1}{2} }=\pm1$, hence $x^s = x\cdot x^{\frac{p^m - 1}{2} } \pm x$).

We now analyze the possible sign combinations of $x_1^s$ and $x_2^s$:
\begin{enumerate}[1)]
	\item If $(x_1^s,x_2^s) = (x_1,x_2)$, substituting $c_1=-c_2$ into the second equation of (1) gives $-c_2x_1+c_2x_2=0$, which simplifies to $x_1 = x_2$. This is a contradiction to $x_1\neq x_2$.
	\item If $(x_1^s,x_2^s) = (x_1, - x_2)$, it then follows from (1) that $x_1 =  - x_2$. When $p^m\equiv1~(\textrm{mod }4)$, $s=\frac{p^m+1}{2}$ is odd. And then, $x_1={x_1^s}=(-x_2)^s=-(x_2)^s=x_2$, which contradicts $x_1 = -x_2 $. Thereby, $\mathcal{C}_p(0,s)$ has no codeword with Hamming weight 2, i.e., $d \ge 3$. When $p^m\equiv3~(\textrm{mod }4)$, for any $\alpha \in \mathbb{F}_{p^m}^*$, $(x_1,x_2)=(\alpha,-\alpha)$ is a solution of \eqref{eq:3.1}. Therefore, $\mathcal{C}_p(0,s)$ contains a codeword with $d=2$. \qed
\end{enumerate}

\dse{3.1~~The first class of optimal $p$-ary cyclic codes ${\mathcal{C}_p}(0,s,t)$ with minimum distance 4}

Let $t =\frac{p^m - 1}{2}+ p^r + 1$ with $0 \le r \le m-1$. In this subsection, we prove that the $p$-ary cyclic code $\mathcal{C}_p(0,s,t)$ is optimal, with parameters $[p^m-1, p^m - 2m - 2,4]$, when $m\equiv 0~(\bmod~4)$.\\

\noindent\textbf{Lemma 3.3.} Let $p \ge 5$ be an odd prime and $m > 2$ be an even integer. Let $t =\frac{p^m - 1}{2}+p^r + 1$ and $s = \frac{p^m + 1}{2}$. If $r=0$ or $\gcd (r,m) = 1$, then $t \notin {C_s}$ and $|{C_t}| = m$.\\

\noindent\textbf{Proof.} We first prove $t \notin {C_s}$. Suppose, for contradiction, that $t \in C_s$. By the definition of cyclotomic cosets, there exists some integer $j$ with $1 \le j \le m-1$ such that $\frac{p^m - 1}{2} + p^r + 1 \equiv s \cdot p^j~(\bmod~p^m - 1)$. Recall $s=\frac{p^m+1}{2}$; substituting this into the congruence gives $\frac{p^m - 1}{2} + p^r + 1 \equiv \frac{p^m+1}{2} \cdot p^j~(\bmod~p^m - 1)$. Rearranging terms, we find 
$p^m-1$ divides the difference $\frac{p^m +1}{2}\cdot p^j-(\frac{p^m- 1}{2}+p^r+1)$. Simplify the difference by expanding and grouping terms involving $p^m$: $\frac{p^m +1}{2} \cdot p^j -\frac{p^m- 1}{2} - p^r - 1= (p^m - 1)(\frac{p^j -1 }{2}) + p^j - p^r- 1$. Since $(p^m - 1)\mid (p^m - 1)\frac{p^j -1 }{2}$, it must divide the remaining term $p^j- p^r - 1$, i.e., $(p^m- 1)\mid (p^j- p^r - 1)$. We analyze this divisibility by considering cases for $r$:
\begin{enumerate}[]
	\item \textbf{Case 1:} $r=0$. Then $p^j- p^0 - 1=p^j-2$, so $(p^m - 1)\mid(p^j-2)$. But $1 \le j \le m-1$ implies $p^j-2\leq p^{m-1}-2$, while $p^m-1>p^{m-1}-2$ (since $p\geq 5$ and $m>2$). This is a contradiction. 
	\item \textbf{Case 2:} $r\neq0$.  When $r=j$, we have ${p^m} - 1$ divides $-1$. This is clearly impossible. When $r > j$, we have ${p^j} - {p^r} - 1 = -({p^j}({p^{r - j}} - 1) + 1)$. Furthermore, we deduce that $p - 1$ divides 1, which is impossible. When $r < j$, we have ${p^j} - {p^r} - 1= {p^r}({p^{j - r}} - 1) - 1$. Then, we get $p - 1$ divides $- 1$, which is also impossible.
\end{enumerate} 

All cases lead to contradictions, so $t \notin {C_s}$. Next, we prove $|{C_t}| = m$, considering the two conditions on $r$:
\begin{enumerate}[]
	\item \textbf{Subcase A :} $r=0$. Then $t=\frac{p^m - 1}{2}+2$. Since $m$ is even and $p$ is odd, $p^m\equiv 1~(\bmod~8)$, so $\frac{p^m - 1}{2}\equiv 0~(\bmod~4)$, hence $t\equiv 0+2~(\bmod~4)$. This means $2\mid t$ and $4\nmid t$. Note also that $4\mid (p^m-1)$. We compute $\gcd(t,p^m-1)$ using properties of the greatest common divisor:
	\begin{align*}
		\gcd(t,p^m-1)&=\frac{1}{2} \gcd(2t,p^m-1)\\
		&=\frac{1}{2}\gcd(p^m-1+4,p^m-1)\\
		&=\frac{1}{2}\gcd(4,p^m-1)\\
		&=2.
	\end{align*}
  Since $0< \gcd(t,p^m-1)=2< p$ (as $p\geq5$), Lemma 2.2 implies $|{C_t}| = m$.
	\item \textbf{Subcase B:} $\gcd (r,m) = 1$. By Lemma 2.2, it suffices to show $\gcd (t,p^m - 1)\gcd (p^h- 1,p^m- 1) \not \equiv 0~(\bmod~p^m - 1)$ for all $0< h < m $. First, compute an upper bound for $\gcd (t,p^m - 1)$.  Since $p^m-1=2\cdot \frac{p^m-1}{2}$, we use the gcd linearity property:
	\begin{align*}
		\gcd(t,p^m-1)&=\gcd{\big(}\frac{{p^m} - 1}{2}+ {p^r} + 1,2\cdot\frac{p^m-1}{2}{\big)}\\
		&\le 2\gcd(\frac{{p^m} - 1}{2}+ {p^r} + 1,\frac{p^m-1}{2})\\
		&=2\gcd({p^r} + 1,\frac{p^m-1}{2})\\
		&\le2\gcd({p^r} + 1,p^m-1)\\
		&=2(p^{\gcd(r,m)}+1)\\
		&=2(p+1).
	\end{align*}
	Next, compute an upper bound for $\gcd(p^h-1,p^m-1)$. By Lemma 2.1, this gcd equals $p^{\gcd(h,m)}-1$. Since $m$ is even and $h<m$, we have $\gcd(h,m)\leq \frac{m}{2}$. Hence, $\gcd(p^h-1,p^m-1)\leq p^{\frac{m}{2}}-1$. Now, multiply the two upper bounds and show the product is less than $p^m-1$ (for $m>2$):
	\begin{align*}
		\gcd (t,{p^m} - 1)\gcd ({p^h} - 1,{p^m} - 1)& \le 2(p+1)(p^{\frac{m}{2}} - 1)\\
		&=2(p^{\frac{m+2}{2}}+p^{\frac{m}{2}}-p-1)\\
		&< {p^m} - 1.
	\end{align*}
	Thus, $\gcd (t,p^m - 1)\gcd (p^h- 1,p^m - 1) \not \equiv 0~(\bmod~p^m- 1)$, which implies the product is not divisible by $p^m - 1$. By Lemma 2.2, $|{C_t}| = m$.
\end{enumerate} 

Combining the above results, we conclude $t \notin {C_s}$ and  $|{C_t}| = m$. \qed\\

\noindent\textbf{Theorem 3.4.} Let $t = \frac{p^m - 1}{2} + p^r + 1$ where $0 \le r \le m-1$ and $m\equiv0~(\bmod~4)$. If $r=0$ or $\gcd(r,m) = 1$, then the $p$-ary cyclic code ${\mathcal{C}_p}(0,s,t)$ has parameters $[p^m - 1,p^m- 2m - 2,4]$ and is optimal.\\

\noindent\textbf{Proof.} By Lemma 3.3, we have $t \notin {C_s}$ and $|{C_t}| = m$. Since the dimension of a cyclic code $\mathcal{C}_p(0,s,t)$ is given by $n-|C_s|-|C_t|$ (where $n=p^m-1$ is the code length), substituting $|C_s|=|C_t|=m$ yields: $\dim({\mathcal{C}_p}(0,s,t))={p^m} - 2m - 2$. Next, we determine the minimum Hamming distance $d$ of $\mathcal{C}_p(0,s,t)$. Since $m\equiv 0~(\bmod~4)$, we have $p^m\equiv1~(\bmod~4)$.
By Lemma 3.2, ${\mathcal{C}_p}(0,s)$ has $d\geq3$, so $\mathcal{C}_p(0,s,t)$ (as a subcode) cannot contain codewords of weight $1$ or $2$. To confirm $d=4$, it suffices to show $\mathcal{C}_p(0,s,t)$ contains no codewords of weight $3$. Suppose, for contradiction, that $\mathcal{C}_p(0,s,t)$ has a codeword of weight $3$. Then there exist scalars ${c_1},{c_2},{c_3}\in\mathbb{F}_p^*$ and distinct elements ${x_1},{x_2},{x_3}\in\mathbb{F}_{{p^m}}^*$ such that:
\begin{align*}
\left\{ \begin{array}{l}
{c_1} + {c_2} + {c_3} = 0,\\
{c_1}x_1^s + {c_2}x_2^s + {c_3}x_3^s = 0,\\
{c_1}x_1^t + {c_2}x_2^t + {c_3}x_3^t = 0.
\end{array} \right.
\end{align*}
Let ${d_2} =\frac{{c_2}}{{c_1}},{d_3} = \frac{{c_3}}{{c_1}}$ (since $c_1\in\mathbb{F}_p^*$, these are well-defined). Dividing the system by $c_1$ gives:
\begin{align}\label{eq:3.2}
\left\{ \begin{array}{l}
1 + {d_2} + {d_3} = 0,\\
x_1^s + {d_2}x_2^s + {d_3}x_3^s = 0,\\
x_1^t + {d_2}x_2^t + {d_3}x_3^t = 0.
\end{array} \right.
\end{align}
Recall $s=\frac{p^m+1}{2}$ and $t=\frac{p^m-1}{2}+p^r+1=s+p^r$. For any $x\in\mathbb{F}_p^*$, $x^{p^m-1}=1$, so $x^s=x\cdot x^{\frac{p^m-1}{2}}=\mu(x)x$, where $\mu(x)=\pm 1$ (the quadratic character of $x$). Similarly, $x^t=x^{s+p^r}=x^s\cdot x^{p^r}=\mu(x)x^{p^r+1}$. By symmetry, we only need to analyze the two non-trivial cases for  $(\mu (x_1),\mu (x_2),\mu (x_3)) $:

\textbf{Case (i):} $(\mu ({x_1}),\mu ({x_2}),\mu ({x_3})) = (1,1,1)$. Here $x_i^s=x_i$ and $x_i^t=x_i^{p^r+1}$ for $i=1,2,3$. From the second equation of \eqref{eq:3.2}: ${x_1} =  - ({d_2}{x_2} + {d_3}{x_3})$. Substitute ${x_1} =-{d_2}{x_2}-{d_3}{x_3}$ into the third equation of \eqref{eq:3.2}:
$${(-{d_2}{x_2}-{d_3}{x_3})^{{p^r} + 1}} + {d_2}x_2^{{p^r} + 1} + {d_3}x_3^{{p^r} + 1} = 0.$$ 
After simplification, we have
$$(d_2^2 + {d_2})x_2^{{p^r} + 1} + {d_2}{d_3}(x_2^{{p^r}}{x_3} + {x_2}x_3^{{p^r}}) + (d_3^2 + {d_3})x_3^{{p^r} + 1} = 0.$$
From the first equation of (2), ${d_3} =  - 1 - {d_2}$. Substitute ${d_3} =  - 1 - {d_2}$ into the coefficients leads $d_2^2 + {d_2} = d_3^2 + {d_3} =  - {d_2}{d_3} \ne 0$ (since $d_2,d_3\in\mathbb{F}_p^*$). Substituting these into the equation gives: 
$$-d_2d_3(x_2^{p^r + 1} -x_2^{p^r}x_3 -x_2x_3^{p^r} +x_3^{p^r + 1}) = 0.$$
Since $- {d_2}{d_3} \ne 0$, the bracket term must be zero: $$x_2^{p^r + 1} -x_2^{p^r}x_3-x_2x_3^{p^r} +x_3^{p^r + 1}=0.$$
Factor the left-hand side: $(x_2^{p^r}-x_3^{p^r})(x_2-x_3)=(x_2 - x_3)^{{p^r} + 1}= 0$. This implies $x_2 - x_3=0$, so ${x_2} = {x_3}$. This is a contradiction with ${x_2} \ne {x_3}$.

\textbf{Case (ii):} $(\mu ({x_1}),\mu ({x_2}),\mu ({x_3})) = (1,1,-1)$. Here $x_1^s=x_1$, $x_2^s=x_2$ $x_3^s=-x_3$ and $x_2^t=x_2^{p^r+1}$,$x_3^t=-x_3^{p^r+1}$. From the second equation of \eqref{eq:3.2}, we obtain $${x_1} =  - {d_2}{x_2} + {d_3}{x_3}.$$ Substituting this into the third equation of \eqref{eq:3.2} yields:
$${(- {d_2}{x_2}+{d_3}{x_3} )^{{p^r} + 1}} + {d_2}x_2^{{p^r} + 1} - {d_3}x_3^{{p^r} + 1} = 0.$$ Expanding and simplifying, we obtain:
$$(d_2^2 + {d_2})x_2^{{p^r} + 1} - {d_2}{d_3}(x_2^{{p^r}}{x_3} + {x_2}x_3^{{p^r}}) + (d_3^2 - {d_3})x_3^{{p^r} + 1} = 0.$$
Using the relation $d_3=-1-d_2$, we simplify the coefficients as follows: $$d_2^2 + {d_2} =  - {d_2}{d_3}, ~ d_3^2 - {d_3} =  - {d_2}{d_3} - 2{d_3}.$$ Substituting these into the above equation gives:
$$ - {d_2}{d_3}x_2^{{p^r} + 1} - {d_2}{d_3}(x_2^{{p^r}}{x_3} + {x_2}x_3^{{p^r}}) - {d_2}{d_3}x_3^{{p^r} + 1} - 2{d_3}x_3^{{p^r} + 1} = 0.$$ 
Divide both sides by $-d_2d_3x_3^{p^r+1}$ (note that  $d_2, d_3, x_3 \neq 0$), we obtain: 
$$ (\frac{x_2}{x_3})^{p^r + 1} - (\frac{x_2}{x_3})^{p^r} +\frac{x_2}{x_3}+ 1=-\frac{2}{d_2}.$$ 
This can be rewritten as: 
$$(\frac{{x_2} + {x_3}}{{x_3}})^{p^r+1} =  - \frac{2}{{d_2}}.$$ 
Let $\theta  = \frac{{x_2} + {x_3}}{{x_3}}$. Then: $$\theta ^{(p^r + 1)(p - 1)} =(- \frac{2}{d_2})^{p-1}=1,$$ and ${x_2} = (\theta  - 1){x_3}$, where $\theta  \ne 1$.

If $r=0$, then $\gcd(({p^r} + 1)(p - 1),{p^m} - 1) = 2(p - 1) $, so ${\theta^{2(p-1)}} = 1$. Hence, ${\theta ^{{p^2} - 1}} = 1$.

If $\gcd (r,m) = 1$ and $m$ is even, then $\frac{{m}}{{\gcd (m,r)}}$ must be even. By Lemma 2.1, we have:
$$\gcd (({p^r} + 1)(p - 1),{p^m} - 1) = ({p^{\gcd (m,r)}} + 1)(p - 1) = {p^2} - 1.$$ 
Since ${\theta ^{({p^r} + 1)(p - 1)}} = 1$, it follows that ${\theta ^{{p^2} - 1}} = 1$. 

Therefore, $\theta \in \mathbb{F}_{p^2} \setminus \{0, 1\}$ and $\theta - 1 \in \mathbb{F}_{p^2}^*$.
Given that $m \equiv 0 \pmod{4}$, we have $2(p^2 - 1) \mid (p^m - 1)$. Then, $(\theta  - 1)^s=(\theta  - 1)^{\frac{p^m - 1}{2(p^2- 1)}(p^2 - 1) + 1}=\theta  - 1$. Now observe:
$${x_2} = x_2^s = {(\theta  - 1)^s}x_3^s = (\theta  - 1)x_3^s =(\frac{{x_2}}{{x_3}})x_3^s = (\frac{{x_2}}{{x_3}})( - {x_3}) =  - {x_2},$$ which implies ${x_2} = 0$. This contradicts the fact that ${x_2} \in \mathbb{F}_{{p^m}}^*$.

From the above analysis, we conclude that the code $\mathcal{C}_p(0,s,t)$ has  the minimum Hamming distance $d=4$. \qed\\

\noindent\textbf{Remark 1.} In 2025, Fan and Zeng [20] proved that for $m\equiv 0~(\bmod~4)$ and $r$ is coprime to $m$, the cyclic code ${\mathcal{C}_5}(0,s,t)$ has parameters $[5^m-1,5^m-2m-2,4]$, where $t = \frac{{5^m} - 1}{2} + {5^r} + 1$ and $s = \frac{{{5^m} + 1}}{{2}}$. This result is the special case of Theorem 3.4 when $p = 5$.\\

\noindent\textbf{Example 3.5.} Let $p = 7$, $m = 4$ and $r=1$. So, $t=\frac{7^4-1}{2}+7^1+1=1208$ and $s=\frac{7^4+1}{2}=1201$. 
 Let $\omega$ be a generator of the multiplicative group $\mathbb{F}_{{7^4}}^*$ with minimal polynomial $\omega^4+5\omega^2+4\omega+3=0$. The resulting cyclic code $C_7(0,1201,1208)$ has parameters $[2400,2391,4]$, and its generator polynomial is $x^9+3x^8+5x^7+5x^6+2x^5+5x^4+2x^3+5x^2+6x+1$.\\

\dse{3.2~~The second class of optimal $p$-ary cyclic codes ${\mathcal{C}_p}(0,s,e)$ with minimum distance 4}

This subsection examines integers $t$ that satisfy the following congruence relation:
\begin{align}\label{eq:3.3}
(p^m-2)t \equiv p^r~(\bmod~p^m-1),
\end{align}
where $m \equiv 0~(\bmod~4)$ and $0\le r \le m-1$.\\ 

\noindent\textbf{Lemma 3.6.} Let $t$ be defined as in \eqref{eq:3.3}, let $s= \frac{{p^m} + 1}{2}$, and let $m$ be an even positive integer. Then $|{C_t}| = m$ and $t \notin {C_s}$.\\

\noindent\textbf{Proof.}  First, since $\gcd(t,p^m-1)=\gcd((p^m-2)t,p^m-1)=\gcd(p^r,p^m-1)=1$, it follows from Lemma 2.2 that $|{C_t}| = m$. Now suppose, for contradiction, that $t \in {C_s}$. Then there exists an integer $j$ with $1\le j \le m-1$ such that $$t\equiv s\cdot p^j~(\bmod~p^m-1).$$ This implies $$(p^m-1)\mid(t-\frac{p^m+1}{2}\cdot p^j).$$ Note that $$t-\frac{p^m+1}{2}\cdot p^j=t-p^j-\frac{p^m-1}{2}\cdot p^j,$$ so we conclude that $$\frac{p^m-1}{2}\mid (t-p^j).$$ On the other hand, from the identity $(p^m-2)t\equiv p^r~(\bmod~p^m-1)$, we have $$(p^m-1)\mid(t+ p^r),$$ and hence $$\frac{p^m-1}{2}\mid (t+p^r).$$ Combining $\frac{p^m-1}{2}\mid (t+p^r)$ and $\frac{p^m-1}{2}\mid (t-p^j)$, we obtain $$\frac{p^m-1}{2}\mid ((t+ p^r)-(t-p^j))=p^r+p^j.$$ That is, $$p^r+p^j \equiv 0 ~(\bmod~\frac{p^m-1}{2}).$$  Since $m$ is even, $\frac{p^m-1}{2}$ is divisible by $p-1$. Therefore, $$p^r+p^j \equiv 0 ~(\bmod~p-1).$$ 
Note that $p^r \equiv 1 \pmod{p - 1}$ and $p^j \equiv 1 \pmod{p - 1}$, so $$1+1 \equiv 0 ~(\bmod~p-1),$$ which implies $p - 1 \mid 2$. For $p\geq 5$, this is impossible. This contradiction implies our initial assumption $t \in {C_s}$ is false, so $t \notin {C_s}$. \qed\\

\noindent\textbf{Theorem 3.7.} Let $t$ be defined as in \eqref{eq:3.3}, let $s= \frac{{p^m} + 1}{2}$, and suppose $m\equiv 0~(\bmod~4)$. Then the $p$-ary cyclic code ${\mathcal{C}_p}(0,s,t)$ has parameters $[{p^m} - 1,{p^m} - 2m - 2,4]$ and is optimal.\\

\noindent\textbf{Proof.} By Lemma 3.6, we have $|{C_t}| = m$ and $t \notin {C_s}$.  For the cyclic code ${\mathcal{C}_p}(0,s,t)$, its length is $n=p^m-1$, and its dimension is determined by the size of the union of cyclotomic cosets $C_0\cup C_s\cup C_t$. Since $C_0=\{0\}$ (size $1$), $|C_s|=m$ (by Lemma 3.1), and $|C_t|=m$ with $C_s\cap C_t=\emptyset$, the dimension is $\dim({\mathcal{C}_p}(0,s,t))=n-|C_0\cup C_s\cup C_t|={p^m} - 2m - 2$. 

Next, we determine the minimum Hamming distance $d$ of ${\mathcal{C}_p}(0,s,t)$. Since $m\equiv 0~(\bmod~4)$, we have $p^m\equiv 1~(\bmod~4)$. By Lemma 3.2, the code ${\mathcal{C}_p}(0,s)$ has no codewords of weight $1$ or $2$, so its subcode ${\mathcal{C}_p}(0,s,t)$ also contains no such codeword. To confirm $d=4$, we only need to show ${\mathcal{C}_p}(0,s,t)$ has no codewords of weight $3$. Suppose, for contradiction, that ${\mathcal{C}_p}(0,s,t)$ has a codeword of weight $3$. Similar to the proof of Theorem 3.4, we need to show that there exist scalars ${c_1},{c_2},{c_3}\in \mathbb{F}_p^*$ and distinct elements ${x_1},{x_2},{x_3}\in \mathbb{F}_{p^m}^*$ such that the system \eqref{eq:3.2} holds.
Let $x_i=y_i^{p^m-2}$ for $i\in\{1,2,3\}$. Note that $$x_i^{\frac{p^m-1}{2}}=(y_i^{p^m-2})^{\frac{p^m-1}{2}}=(y_i^{\frac{p^m-1}{2}})^{-1},$$ so $\mu(x_i)=\mu(y_i)$. Then \eqref{eq:3.2} becomes
\begin{align}\label{eq:3.4}
\left\{ \begin{array}{l}
1 + {d_2} + {d_3} = 0,\\
y_1^{-s} + {d_2}y_2^{-s} + {d_3}y_3^{-s} = 0,\\
y_1^{p^r} + {d_2}y_2^{p^r} + {d_3}y_3^{p^r} = 0,
\end{array} \right.
\end{align} where $d_2=\frac{c_2}{c_1}$, $d_3=\frac{c_3}{c_1}$.
From the third equation in \eqref{eq:3.4}, we obtain $y_1^{p^r}=(-d_2y_2-d_3y_3)^{p^r}$. It is straightforward to show that 
 $$y_1^{p^r}=(-d_2y_2-d_3y_3)^{p^r},$$
which implies  $$y_1^{-p^r}=(-d_2y_2-d_3y_3)^{-p^r}.$$
By symmetry of $\mu(x_i)$, our analysis can be restricted to just two non-trivial cases of $(\mu(x_1),\mu(x_2),\mu(x_3))$:

\textbf{Case (i):} $(\mu(x_1),\mu(x_2),\mu(x_3))=(1,1,1)$. By $\mu(x_i)=\mu(y_i)$, this implies $(\mu(y_1),\mu(y_2),\mu(y_3))=(1,1,1)$, but the key simplification comes from the second equation of \eqref{eq:3.4}: $$y_1^{-1}=-d_2y_2^{-1}-d_3y_3^{-1}.$$ Raising both sides to the  $p^r$-th power yields
$$y_1^{-p^r}=(-d_2y_2^{-1}-d_3y_3^{-1})^{p^r}.$$
Combining this with the earlier expression for $y_1^{-p^r}$, we get
$$(-d_2y_2^{-1}-d_3y_3^{-1})^{p^r}=(-d_2y_2-d_3y_3)^{-p^r}.$$
Simplifying algebraically, we obtain 
$$d_2d_3y_2^{-p^r}y_3^{p^r}+d_2d_3y_3^{-p^r}y_2^{p^r}+d_2^2+d_3^2-1=0.$$
Let $z=(\frac{y_3}{y_2})^{p^r}$. From $1+d_2+d_3=0$, it follows that $d_2^2+d_3^2-1=-2d_2d_3$. Substituting, the above equation becomes $$z+z^{-1}-2=0,$$ i.e., $z^2-2z+1=0$, so $z=1$. Hence $(\frac{y_3}{y_2})^{p^r}=1$. Since $\gcd(p^r,p^m-1)=1$, we conclude $\frac{y_3}{y_2}=1$. Then $$\frac{x_2}{x_3}=\frac{y_2^{p^m-2}}{y_3^{p^m-2}}=\frac{y_3}{y_2}=1,$$ so $x_2=x_3$, a contradiction.

\textbf{Case (ii):} $(\mu(x_1),\mu(x_2),\mu(x_3))=(1,1,-1)$. By $\mu(x_i)=\mu(y_i)$, this implies $(\mu(y_1),\mu(y_2),\mu(y_3))=(1,1,-1)$. From the second equation of \eqref{eq:3.4}: $$y_1^{-1}=-d_2y_2^{-1}+d_3y_3^{-1}.$$ Applying the $p^r$-th power of both sides gives
$$y_1^{-p^r}=(-d_2y_2^{-1}+d_3y_3^{-1})^{p^r}.$$
A similar algebraic manipulation leads to 
$$d_2d_3y_2^{-p^r}y_3^{p^r}-d_2d_3y_3^{-p^r}y_2^{p^r}+d_2^2-d_3^2-1=0.$$
Again, let $z=(\frac{y_3}{y_2})^{p^r}$. From $1+d_2+d_3=0$, we get $d_2^2-d_3^2-1=-2d_3$. Then the above equation becomes $$d_2z^2-2z-d_2=0.$$
Since ${d_2}\neq 0$, by Lemma 2.5,  there exists a solution $z\in\mathbb{F}_{p^2}^*$. If $m\equiv0~(\bmod~4)$, then by Lemma 2.6, $\mu(z)=1$. However, 
$$\mu(z)=\mu((\frac{y_3}{y_2})^{p^r})=\mu((\frac{x_2}{x_3})^{p^r})=(\frac{x_2}{x_3})^{\frac{p^m-1}{2}\cdot{p^r}}=-1,$$ a contradiction.

Therefore, ${\mathcal{C}_p}(0,s,t)$ has no codeword of Hamming weight 3.  We conclude that the minimum distance is at least 4. \qed\\

\noindent\textbf{Remark 2.} It can be readily verified through substituting $t=p^m-2$ into congruence \eqref{eq:3.3} that $t=p^m-2$ is a solution. Through direct computation based on the properties of cyclotomic cosets, we further show that all solutions to congruence \eqref{eq:3.3} lie in the cyclotomic coset $C_{p^m-2}$. Consequently, the cyclic code ${\mathcal{C}_p}(0,s,t)$ in Theorem 3.7 is equivalent to ${\mathcal{C}_p}(0,s,p^m-2)$, where $s=\frac{p^m+1}{2}$.\\

\noindent\textbf{Example 3.8.} Let $p = 5$, $m = 4$ and $r=1$. Then, $s=\frac{5^4+1}{2}=313$ and $t =619$ is a solution of \eqref{eq:3.3}. Let $\omega $ be a generator of the multiplicative group $\mathbb{F}_{{5^4}}^*$ with minimal polynomial $\omega^4+4\omega^2+4\omega+2=0$. The resulting cyclic code $\mathcal{C}_5(0,313,619)$ has parameters $[624,615,4]$, and its generator polynomial is ${x^9}+{x^8}+4{x^7} + 3{x^6} + {x^5} + {x^4} +2{x^2} +3{x}+ 4$.\\

\noindent\textbf{Example 3.9.} Let $p = 7$, $m = 4$ and $r=2$. Then, $s=\frac{7^4+1}{2}=1201$ and $t =2351$ is a solution to \eqref{eq:3.3}. Let $\omega $ be a generator of the multiplicative group $\mathbb{F}_{{7^4}}^*$ whose minimal polynomial $\omega^4+5\omega^2+4\omega+3=0$. The resulting cyclic code $\mathcal{C}_5(0,1201,2351)$ has parameters $[2400,2391,4]$, and its generator polynomial is ${x^9}+5{x^8}+3{x^7} + 3{x^6} + 6{x^5} + 5{x^4} +6{x^2} + 6$.\\

\dse{3.3~~The third class of optimal $p$-ary cyclic codes ${\mathcal{C}_p}(0,s,t)$ with minimum distance 4}

Throughout this subsection, we let $\zeta$ denote a primitive fourth root of unity in the finite field $\mathbb{F}_{p^m}$. For any nonzero elements $y_1,y_2 \in \mathbb{F}_{p^m}^*$, we denote their quartic character pair by $(\eta(y_1),\eta(y_2))$, where $\eta:\mathbb{F}_{p^m}^* \Longrightarrow \{1,\zeta,\zeta^2,\zeta^3\}$ is the quartic character of $\mathbb{F}_{p^m}^*$ (i.e., $\eta(y)=y^{\frac{p^m-1}{4}}$ for all $y\in \mathbb{F}_{p^m}^*$). By exploiting the algebraic properties of quartic characters, we establish the optimality of ${\mathcal{C}_p}(0,s,t)$, where $s = \frac{{{p^m} + 1}}{{2}}$.\\

\noindent\textbf{Lemma 3.10.} Let $t=\frac{r(p^m-1)}{4}-1$ with $ r \in \{1, 3\} $, where $p \ge 5$ is an odd prime and $m\equiv 0~(\bmod~4)$. Then $|{C_t}| = m$ and $t \notin {C_s}$.\\

\noindent\textbf{Proof.} Since $t=\frac{r(p^m-1)}{4}-1$ and $m\equiv 0~(\bmod~4)$, it follows that $t$ is odd and $\gcd(t,p^m-1)=1$. By Lemma 2.2, we conclude that $|{C_t}| = m$.

Now suppose, for contradiction, that $t \in {C_s}$. Then there exists an integer $i$ with $1\le i\le m-1$ such that $$t\equiv sp^i~(\bmod~p^m-1),$$ i.e., $$\frac{r(p^m-1)}{4}-1\equiv sp^i~(\bmod~p^m-1).$$ This implies  $$(p^m-1)\mid (\frac{p^m+1}{2}p^i-\frac{r(p^m-1)}{4}+1).$$ Note that $$\frac{p^m+1}{2}p^i-\frac{r(p^m-1)}{4}+1=\frac{p^m-1}{4}(2p^i-r)+p^i+1.$$ Therefore, $$\frac{p^m-1}{4}\mid(p^i+1).$$ However, for $1\le i\le m-1$, we have $p^i+1\leq p^{m-1}+1<\frac{p^m-1}{4}$, since $\frac{p^m-1}{p-1}>p^{m-1}+1$ for $p\geq5$ and $m\geq4$. This is a contradiction. Hence, $t\notin {C_s}$.\qed\\

\noindent\textbf{Theorem 3.11.} Let $p \ge 5$ be an odd prime and $m$ a positive integer. Define $$t=\frac{r(p^m-1)}{4}-1, \textrm{ with } r \in \{1, 3\},  \textrm{ and } s = \frac{{{p^m} + 1}}{{2}}.$$ Then, the cyclic code ${\mathcal{C}_p}(0,s,t)$ achieves optimal parameters $[{p^m} - 1,{p^m} - 2m - 2,4]$  if and only if one of the following congruence conditions holds:
\begin{enumerate}[1.]
	\item  $p\equiv 1~(\bmod~4)$ and $m\equiv 0~(\bmod~4)$. 
	\item $p\equiv 3~(\bmod~4)$ and $m\equiv 0~(\bmod~8)$.
\end{enumerate} 

\noindent\textbf{Proof.} We restrict our attention to the case $r=3$, as the case $r=1$ can be proved in a similar manner.

From Lemma 3.10, we have $|{C_t}| = m$ and $t \notin {C_s}$, which implies that the dimension of ${\mathcal{C}_p}(0,s,t)$ is $p^m-2m-2$. Given that $m\equiv 0~(\bmod~4)$, it follows that $p^m\equiv1~(\bmod~4)$. By Lemma 3.2, the code ${\mathcal{C}_p}(0,s,t)$ contains no codewords of Hamming weight 1 or 2. We now prove that it also contains no codewords of Hamming weight $3$. Suppose, to the contrary, that there exists a nonzero codeword of weight $3$. Then there exist nonzero scalars ${c_1},{c_2},{c_3}\in \mathbb{F}_p^*$ and distinct elements ${x_1},{x_2},{x_3}\in \mathbb{F}_{p^m}^*$ such that
\begin{align*}
\left\{ \begin{array}{l}
{c_1} + {c_2} + {c_3} = 0,\\
{c_1}x_1^s + {c_2}x_2^s + {c_3}x_3^s = 0,\\
{c_1}x_1^t + {c_2}x_2^t + {c_3}x_3^t = 0.
\end{array} \right.
\end{align*}
Define $${d_1} =\frac{c_1}{c_3},{d_2} = \frac{c_2}{c_3},  {y_1} =\frac {x_1}{x_3},{y_2} = \frac{x_2}{x_3},$$ where $y_1,y_2\neq0,1$. Then $  $the system becomes
\begin{align}\label{eq:3.5}
\left\{ \begin{array}{l}
{d_1} + {d_2} + 1= 0,\\
{d_1}y_1^s + {d_2}y_2^s + 1 = 0,\\
{d_1}y_1^t + {d_2}y_2^t + 1 = 0.
\end{array} \right.
\end{align}\\
By symmetry, it suffices to consider the following cases for the pair $(\mu(y_1),\mu(y_2))$: $$(1,1), (1,-1), (-1,-1).$$ We provide a detailed analysis only for the case $(-1,-1)$, as the other cases can be handled similarly.

Assume $\mu(y_1)=\mu(y_2))=-1$. From the second equation in \eqref{eq:3.5}, we derive $$y_2=d_2^{-1}(1-d_1y_1),$$ and consequently, $$y_2^{-1}=d_2(1-d_1y_1)^{-1}.$$ We now consider four subcases based on the values of $\eta(y_1)$ and $\eta(y_2)$:

(a)  $(\eta(y_1),\eta(y_2))=(\zeta,\zeta)$. The third equation in \eqref{eq:3.5} becomes
$$-\zeta d_1y_1^{-1}-\zeta d_2y_2^{-1}+1=0.$$
Substituting  $y_2^{-1}=d_2(1-d_1y_1)^{-1}$, we obtain
$$-\zeta  d_1y_1^{-1}-\zeta d_2^2(1-d_1y_1)^{-1}+1=0.$$
Multiplying through by $y_1(1-d_1y_1)$ yields
$$-\zeta  d_1(1-d_1y_1)-\zeta d_2^2y_1+y_1(1-d_1y_1)=0,$$
which simplifies to the quadratic equation
$$-d_1y_1^{2}+(-2\zeta d_1-\zeta+1)y_1-\zeta d_1=0.$$

(b) $(\eta(y_1),\eta(y_2))=(\zeta,-\zeta)$. The third equation becomes
$$-\zeta d_1y_1^{-1}+\zeta d_2y_2^{-1}+1=0.$$
Substituting for $y_2^{-1}$ and simplifying leads to
$$-d_1y_1^{2}+(2\zeta d_1^2+2\zeta d_1+\zeta+1)y_1-\zeta d_1=0.$$

(c) $(\eta(y_1),\eta(y_2))=(-\zeta,\zeta)$. The third equation becomes
$$\zeta d_1y_1^{-1}-\zeta d_2y_2^{-1}+1=0,$$
which reduces to
$$-d_1y_1^{2}+(-2\zeta d_1^2-2\zeta d_1-\zeta+1)y_1+\zeta d_1=0.$$

(d) $(\eta(y_1),\eta(y_2))=(-\zeta,-\zeta)$. The third equation becomes
$$\zeta d_1y_1^{-1}+\zeta d_2y_2^{-1}+1=0,$$ yielding 
$$-d_1y_1^{2}+(2\zeta d_1+\zeta+1)y_1+\zeta d_1=0.$$

Clearly, $d_1\neq 0$. We now consider the two cases separately: 
\begin{itemize}
	\item [1)]If $p\equiv1~(\bmod~4)$, then $\zeta \in \mathbb{F}_p$. By Lemma 2.5, all solutions $y_1$ lie in $\mathbb{F}_{p^2}^*$. If $m\equiv 0~(\bmod~4)$, then Lemma 2.6 implies $\mu(y_1)=1$, contradicting the assumption that $\mu(y_1)=-1$. 
	\item [2)]If $p\equiv3~(\bmod~4)$, then $\zeta \in \mathbb{F}_{p^2}$. By Lemma 2.5, all solutions $y_1$ lie in $\mathbb{F}_{p^4}^*$. If $m\equiv 0~(\bmod~8)$, then Lemma 2.6 implies $\mu(y_1)=1$, again contradicting the assumption that $\mu(y_1)=-1$.
\end{itemize}

In both cases, the assumption leads to a contradiction. Therefore, no codeword of weight $3$ exists in ${\mathcal{C}_p}(0,s,t)$. Since it was previously established that there are no codewords of weight $1$ or $2$, and the Singleton bound implies that the minimum distance cannot exceed $4$, we conclude that the code has minimum distance exactly $4$. Hence, ${\mathcal{C}_p}(0,s,t)$ is optimal under the stated conditions. \qed\\

\noindent\textbf{Remark 3.} It is straightforward to verify that 
 $t_1=\frac{p^m-1}{4}-1$ and $t_2=\frac{3(p^m-1)}{4}-1$ belong to distinct cyclotomic cosets modulo $p^m-1$. Indeed, suppose for contradiction that there exists an integer 
$j$ with $1\le j\le m-1$ such that $$\frac{p^m-1}{4}-1\equiv(\frac{3(p^m-1)}{4}-1)\cdot p^j~(\bmod~p^m-1).$$ Then,
\[(p^m-1)\mid \Big((\frac{p^m-1}{4}-1)(p^j-1)+\frac{p^j-1}{2}(p^m-1)+\frac{p^m-1}{2}\Big),\]
which simplifies to $$(p^m-1)\mid((\frac{p^m-1}{4}-1)(p^j-1)+\frac{p^m-1}{2}).$$ This implies $$\frac{p^m-1}{4}\mid (\frac{p^m-1}{4}-1)(p^j-1).$$ Since $\gcd(\frac{p^m-1}{4},\frac{p^m-1}{4}-1)=1$, it follows that $$\frac{p^m-1}{4}\mid(p^j-1).$$ However, observe that $$\frac{p^m-1}{4}> p^{m-1}+1>p^j-1 \text{ for all } 1\leq j\leq m-1,$$ which contradicts the divisibility condition. Therefore, no such $j$ exists, and hence $t_1$ and $t_2$ belong to distinct cyclotomic cosets. Consequently, the cyclic codes ${\mathcal{C}_p}(0,s,t_1)$ and ${\mathcal{C}_p}(0,s,t_2)$ are different.\\

\noindent\textbf{Example 3.12.} Let $p = 5$, $m = 4$ and $r=1$, so that $t =\frac{5^4-1}{4}-1=155$ and $s=\frac{5^4+1}{2}=313$. Let $\omega $ be the generator of $\mathbb{F}_{{5^4}}^*$ with $\omega^4+4\omega^2+4\omega+2=0$. Then the resulting cyclic code $\mathcal{C}_5(0,313,155)$ has parameters $[624,615,4]$, and its generator polynomial is ${x^9}+3{x^8}+3{x^7} + 4{x^5} + 2{x^3} +3{x}+ 4$.\\

\noindent\textbf{Example 3.13.} Let $p = 5$, $m = 4$ and $r=3$, so that $t =\frac{3\cdot(5^4-1)}{4}-1=467$ and $s=\frac{5^4+1}{2}=313$. Let $\omega $ be the generator of $\mathbb{F}_{{5^4}}^*$ with $\omega^4+4\omega^2+4\omega+2=0$. Then the resulting cyclic code $\mathcal{C}_5(0,313,467)$ has parameters $[624,615,4]$, and its generator polynomial is ${x^9}+{x^7} + 3{x^6} + 3{x^5} + 2{x^4} +3{x^3} +3{x}+ 4$.\\

\noindent\textbf{Remark 4.} When ${p^m} \equiv 1~(\bmod~4)$, the integer $s=\frac{p^m + 1}{2}$ is odd. Moreover, $$\gcd (s,{p^m} - 1) = \gcd (\frac{p^m+1}{2},{p^m} - 1) = 1.$$ Note that  $${s^2}= {(\frac{p^m + 1}{2})^2}= \frac{p^m - 1}{4}({p^m} - 1) + ({p^m} - 1) + 1,$$ which implies ${s^2} \equiv 1~(\bmod~{p^m} - 1)$. Hence, ${s^{ - 1}} \equiv s~(\bmod~ {p^m} - 1)$. Therefore, the cyclic code $\mathcal{C}_p(0,s,t)$ is equivalent to $\mathcal{C}_p(0,1,ts^{-1})$ when ${p^m} \equiv 1~(\bmod~4)$. In this section, the optimal $p$-ary cyclic codes $\mathcal{C}_p(0,s,t)$ constructed above are summarized in Table 1. For comparison, Table 2 lists the currently known $p$-ary cyclic codes $\mathcal{C}_p(0,1,t)$ with parameters $[{p^m}-1,{p^m}-2m-2,4]$. A comparative analysis confirms that the codes $\mathcal{C}_p(0,s,t)$ presented in this work are new. \\

\begin{table}[H]
\caption{Our results $\mathcal{C}_p(0,\frac{p^m+1}{2},t)$ whose parameters are $[{p^m}-1,{p^m}-2m-2,4]$.}
\setlength{\tabcolsep}{2pt}
\small
\begin{tabular*}{14.3cm}{@{\extracolsep{\fill}}l l l l}
\hline
&$t$  &Conditions  &Reference \\
\hline
1) &$t=\frac{p^m-1}{2}+{p^r}+1$  &$m\equiv 0~(\bmod~4)$      &Theorem 3.4\\
& & $r=0$ or $\gcd(r,m)=1$ &\\
2) & $(p^m-2)t\equiv {p^r}~(\bmod~{p^m}-1)$ & $m\equiv 0~(\bmod~4)$ and $0 \le r \le m-1$ & Theorem 3.7\\
3) &$t=\frac{r(p^m-1)}{4}-1$, $r\in \{1,3\}$ & $p\equiv 1~(\bmod~4)$ and $m\equiv 0~(\bmod~4)$ &Theorem 3.11\\
& & or $p\equiv 3~(\bmod~4)$ and $m\equiv 0~(\bmod~8)$ & \\
\bottomrule
\end{tabular*}
\end{table}

\begin{table}[H]
\caption{Known codes $\mathcal{C}_p(0,1,t)$ whose parameters are $[{p^m}-1,{p^m}-2m-2,4]$.}
\setlength{\tabcolsep}{17pt}
\small
\begin{tabular*}{14.3cm}{@{\extracolsep{\fill}}l l l l}
\hline
& $t$ & Conditions & Reference \\
\hline
  1) &$t={p^r}+1$ & $0\leq r \leq m-1$ and $r \ne \frac{m}{2}$ & [6, Theorem 1]\\
  2) &$t={p^m}-2$ & $m> 1$ & [6, Theorem 2]\\
\bottomrule
\end{tabular*}
\end{table}

\dse{4~~Optimal quinary cyclic codes ${\mathcal{C}_5}(0,s,t)$}
We now turn our attention to the construction of optimal quinary cyclic codes. In this section, we establish four distinct families of optimal quinary cyclic codes, denoted ${\mathcal{C}_5}(0,s,t)$, through a systematic investigation of the solution spaces of specific polynomial equations over finite fields. Recall that $\omega $ denotes a multiplicative generator of $\mathbb{F}_{{5^m}}^*$, so $\mathbb{F}_{{5^m}}^*=\langle \omega\rangle$. Then, the subgroup $\mathbb{F}_5^*$ is generated by $\omega^{\frac{5^m-1}{4}}$. In particular, we have $$2 = \omega ^{\frac{5^m - 1}{4}} \text{ and } -2 = \omega ^{3 \cdot \frac{5^m - 1}{4}}.$$ Consequently, the quadratic character $\mu(\pm 2)$ over $\mathbb{F}_5$ satisfies:
\begin{eqnarray*} 
\mu ( \pm 2) = \left\{ \begin{array}{l}
1, ~~~$if$~m~$is$~$even$,\\
-1,~$if$~m~$is$~$odd$.
\end{array} \right. 
\end{eqnarray*}\\
\noindent\textbf{Lemma 4.1.} Let ${C_s}$ be the 5-cyclotomic coset modulo $5^m - 1$ containing $s$, where $s = \frac{5^m + 1}{2}$. Then $t \notin {C_s}$ if any of the following conditions is satisfied:

1) $t$ is even;

2) $t \equiv 3 \pmod{4}$ and $m$ is even;

3) $t \equiv 1~(\bmod~4)$ and $m$ is odd.\\

\noindent\textbf{Proof.} Assume first that $t$ is even. Then for any integer $i$, the product ${5^i}t$ is also even. Since ${5^m} - 1$ is even and $\frac{5^m + 1}{2}$ is odd, it follows that $${5^i}t \not\equiv \frac{5^m + 1}{2}~(\bmod {5^m} - 1).$$ Hence, $t \notin {C_s}$.

Now suppose  $t \equiv 3~(\bmod~4)$ and $m$ is even. Assume, for contradiction, that  $t \in {C_s}$. Then there exists some $i$ with $1 \le i \le m - 1$ such that $$t \equiv s \cdot {5^i}~(\bmod~{5^m} - 1).$$ This implies $$({5^m} - 1)\mid {5^i} \cdot (\frac{5^m- 1}{2}) + {5^i} - t.$$ Since $m$ is even, we have $\frac{5^m- 1}{2}\equiv 0~(\bmod~4)$, and thus ${5^i} - t \equiv 0~(\bmod~4)$, which yields $t \equiv 1~(\bmod~4)$, contradicting the assumption that $t \equiv 3 \pmod{4}$. Therefore, $t \notin {C_s}$ in this case.

Finally, consider the case where $t \equiv 1~(\bmod~4)$ and $m$ is odd. Again, assume $t \in {C_s}$. Then for some $l$ with $1 \le l \le m - 1$, $$({5^m} - 1)\mid {5^l} \cdot (\frac{5^m- 1}{2}) + {5^l} - t.$$ It follows that $${5^l} \cdot (\frac{5^m - 1}{2}) + {5^l} - t\equiv 0~(\bmod~4).$$ Note that since $m$ is odd, $\frac{5^m - 1}{2} \equiv 2~(\bmod~4)$, and since $t \equiv 1~(\bmod~4)$, we have ${5^l} - t \equiv 0~(\bmod~4)$. However, $${5^l} \cdot (\frac{5^m - 1}{2}) + {5^l} - t\equiv 1\cdot2+0 \equiv0~(\bmod~4),$$ which is not divisible by $4$. This contradiction shows that $t\notin {C_s}$.\qed\\

In the theoretical study of quinary cyclic codes, characterizing optimality conditions constitutes a fundamental problem. It was established in Ref. [20] that necessary and sufficient conditions for the optimality of the code $\mathcal{C}_5(0,s,t)$ with $s=\frac{5^m+1}{2}$ are known. Our work builds directly upon this result.\\

\noindent\textbf{Theorem 4.2} ([20, Theorem 1]). Let $m$ be a positive integer and let $t$ satisfy $t \equiv 1,2,\text{or} ~3~(\bmod~4)$. Assume $t \notin {C_s}$ and $|{C_t}| = m$. The quinary cyclic code ${\mathcal{C}_5}(0,s,t)$ has parameters $[{5^m} - 1,{5^m} - 2m - 2,4]$ if and only if the following conditions hold:\\
C1: The equation ${x^t} + {( - 1)^t}{(x + 3)^t} + 3 = 0$ admits only the solution $x=1$ in $\mathbb{F}_{{5^m}}^*$ with $\mu (x) = \mu (x + 3) = 1$;\\
C2: The equation ${x^t} + {(x + 3)^t} + 3 = 0$ admits no solution in $\mathbb{F}_{{5^m}}^*$ with $\mu (x) = 1$ and $\mu (x + 3) =  - 1$; \\
C3: The equation ${x^t} + {( - 1)^t}{(x - 3)^t} + 3 = 0$ admits no solution in $\mathbb{F}_{{5^m}}^*$ with $\mu (x) = \mu (x - 3) =  - 1$.

\dse{4.1~~The first class of optimal quinary cyclic codes ${\mathcal{C}_5}(0,s,t)$ with minimum distance 4}

Suppose that $m$ is odd and $r$ is an integer with $0\le r \le m-1$. In this subsection, we study the quinary cyclic code ${\mathcal{C}_5}(0,s,t)$, where $t$ is a solution of the congruence
\begin{align}\label{eq:4.1}
7t\equiv -2 \cdot {5^r}~(\textrm{mod~} {5^m}-1).
\end{align}

Since $\gcd(5,7)=1$, by Euler's Theorem, we have $5^{\varphi(7)} \equiv 1~(\bmod~7)$, where $\varphi(\cdot)$ denotes Euler's totient function. Note that $\varphi(7)=6$, so $5^6 \equiv 1~(\bmod~7)$, i.e., $7\mid (5^6-1)$. Recall that $$\gcd(5^m-1,5^6-1)=5^{\gcd(m,6)}-1.$$ When $m$ is odd, $\gcd(m,6)\le 3$, which implies $7\nmid 5^{\gcd(m,6)}-1$. Hence, $\gcd(7,5^m-1)=1$ when $m$ is odd. Therefore, the congruence \eqref{eq:4.1} always admits a unique solution modulo $5^m-1$. \\

\noindent\textbf{Theorem 4.3.} Let $t$ be defined as in \eqref{eq:4.1}, $s = \frac{5^m + 1}{2}$ and let $m$ be an odd integer. Then the quinary cyclic code $\mathcal{C}_5(0,s,t)$ has parameters $[5^m - 1, 5^m - 2m - 2, 4]$ and is optimal.\\

\noindent\textbf{Proof.} It is straightforward to verify that $t \equiv 2~(\bmod~4)$. By Lemma 3.1 and Lemma 4.1, we have $|C_s|=m$ and $t \notin {C_s}$. From [22, Lemma 9], it follows directly that $|{C_t}| = m$. Hence, the dimension of $\dim({\mathcal{C}_5}(0,s,t))$ is ${5^m} - 2m - 2$. We now verify the conditions of Theorem 4.2. First, we consider the equation $${x^t} + {(x + 3)^t} + 3 = 0$$ over $\mathbb{F}_{{5^m}}$. Let $y=x+3$. Since $\gcd(7,5^m-1)=1$, there exist $a,b \in \mathbb{F}_{{5^m}}^*$ such that $x = {a^7}$ and $y={b^7}$. Substituting into the equation yields the system:
\begin{align*}
\left\{ \begin{array}{l}
{a^{-2}} + {b^{-2}} + 3= 0,\\
{a^7} - {b^7} + 3 = 0.
\end{array} \right.
\end{align*} 
Multiplying the first equation by $a^2b^2$ gives the equivalent system:
\begin{align}\label{eq:4}
\left\{ \begin{array}{l}
{a^2} + {b^2} + 3{a^2}{b^2} = 0,\\
{a^7} - {b^7} + 3 = 0.
\end{array} \right.
\end{align}
By eliminating the variable $a$, we derive a univariate polynomial equation in $b$: 
\begin{equation*}
	\begin{split}
2b^{28}+3b^{26}+3b^{24}+3b^{21}+2b^{19}+4b^{18}+2b^{17}+b^{16}\\+2b^{12}+b^{11}+2b^{10}+4b^9+4b^7+b^4+4b^2+4=0.
\end{split}
\end{equation*}
 Over $\mathbb{F}_5$, this factors as:
 \begin{equation*}
 	\begin{split}
 	~~~~~~2(b+1)^2({b^6}+{b^5}+{b^4}+3{b^3}+{b^2}+2b+4)~~~~~~~~~~\\
 	({b^{10}}+{b^9}+{b^8}+2{b^7}+{b^6}+3{b^5}+{b^4}+4{b^3}+b+1)~~~~~\\
 	({b^{10}}+{b^9}+2{b^8}+4{b^7}+2{b^6}+4{b^4}+{b^3}+3{b^2}+2b+3)=0.
 	\end{split}
\end{equation*}
 If $b=-1$, then $y=x+3=-1$ and $x=1$. This gives a solution $x=1$ with $\mu (x) = \mu (x + 3) = 1$, but not with $\mu (x + 3) =  - 1$. By Lemma 2.3, the polynomial of degree $6$ factors into three irreducible quadratics over $\mathbb{F}_{5^3}$, and each polynomial of degree $10$ factors into five irreducible quadratics over ${\mathbb{F}_{5^5}}$. Since $\gcd(2,m)=1$, Lemma 2.4 implies that none of these quadratic factors has a root in $\mathbb{F}_{5^m}^*$.

Therefore, the equation ${x^t} + {(x + 3)^t} + 3 = 0$ has exactly one solution $x=1$ in $\mathbb{F}_{5^m}^*$ with $\mu (x) = \mu(x+3)=1$, and no solution with $\mu (x)=1$ and $\mu(x+3)=-1$.

Next, we consider the equation $${x^t} + {(x - 3)^t} + 3 = 0$$ over $\mathbb{F}_{5^m}$. Let $y = x - 3$. Again, write $x = {a^7}$, $y={b^7}$ for some $a,b\in \mathbb{F}_{5^m}^*$. The equation thus simplifies to
\begin{align*}
\left\{ \begin{array}{l}
{a^{-2}} + {b^{-2}} + 3= 0,\\
{a^7} - {b^7} - 3 = 0.
\end{array} \right.
\end{align*} 
Multiplying the first equation by $a^2b^2$ yields the equivalent system:
\begin{align}\label{eq:5}
\left\{ \begin{array}{l}
{a^2} + {b^2} + 3{a^2}{b^2} = 0,\\
{a^7} - {b^7} - 3 = 0.
\end{array} \right.
\end{align}
Eliminating $a$ gives:
\begin{equation*}
	\begin{split}
		2{b^{28}}+3{b^{26}}+3{b^{24}}+2{b^{21}}+3{b^{19}}+4{b^{18}}+3{b^{17}}+~~~\\
		{b^{16}}+2{b^{12}}+4{b^{11}}+2{b^{10}}+{b^9}+{b^7}+{b^4}+4{b^2}+4=0.
	\end{split}
\end{equation*}
 Complete factorization over $\mathbb{F}_5$ yields:
\begin{equation*}
	\begin{split}
	2(b+4)^2({b^6}+4{b^5}+{b^4}+2{b^3}+{b^2}+3b+4)~~~~~~~~~~~~\\
	{b^{10}}+4{b^9}+{b^8}+3{b^7}+{b^6}+2{b^5}+{b^4}+{b^3}+4b+1)~~~~~~~\\
	({b^{10}}+4{b^9}+2{b^8}+{b^7}+2{b^6}+4{b^4}+4{b^3}+3{b^2}+3b+3)=0.
	\end{split}
\end{equation*}
If $b=1$, then $y=1$ and $x=4$, but $\mu(x-3)=\mu (1) = 1$, contradicting the requirement $\mu (x-3) = -1$. The irreducible factors of degrees $6$ and $10$ again have no roots in $\mathbb{F}_{{5^m}}^*$ by Lemma 2.4, since $\gcd(2,m)=1$.

Hence, the equation ${x^t} + {(x - 3)^t} + 3 = 0$ has no solution $x\in\mathbb{F}_{5^m}^* $ with $\mu (x) = \mu(x-3)=-1$.

In conclusion, all conditions of Theorem 4.2 are satisfied for odd $m$, and thus ${\mathcal{C}_5}(0,s,t)$ is optimal.\qed

\dse{4.2~~The second class of optimal quinary cyclic codes ${\mathcal{C}_5}(0,s,t)$ with minimum distance 4}

Let $t = {5^{m - 1}} - 3$, where $m$ is an odd positive integer. In this subsection, we analyze the optimality of the quinary cyclic code ${\mathcal{C}_5}(0,s,t)$.\\

\noindent\textbf{Lemma 4.4.} Let $t = {5^{m - 1}} - 3$ with $m>1$. If $m$ is odd, then $|{C_t}| = m$.\\

\noindent\textbf{Proof.} Since $m$ is odd, we have $\gcd(7,5^m-1)=1$. Then,  
\begin{align*}
\gcd(t,5^m-1)&=\gcd(5t,5^m-1)\\
&=\gcd(5^m-1-14,5^m-1)\\
&=\gcd(14,5^m-1)\\
&\le \gcd(2,5^m-1)\gcd(7,5^m-1)\\
&=2. 
\end{align*}
By Lemma 2.2, it follows that $|{C_t}| = m$. \qed\\

\noindent\textbf{Theorem 4.5.} Let $t = {5^{m - 1}} - 3$, $s = \frac{5^m + 1}{2}$, and suppose $m$ is  an odd positive integer such that $9\nmid m$. Then the quinary cyclic code $\mathcal{C}_5(0,s,t)$ has parameters $[5^m - 1, 5^m - 2m - 2, 4]$ and is optimal.\\

\noindent\textbf{Proof.} It is clear that $t\equiv 2~(\bmod~4)$. From Lemma 4.1, we derive $t \notin {C_s}$, and by Lemma 4.4, $|{C_t}| = m$. Therefore, the dimension of $\dim({\mathcal{C}_5}(0,s,t))$ is ${5^m} - 2m - 2$. We now verify that the three conditions of Theorem 4.2 are satisfied. 

(i) First, we consider the equation $${x^t} + {( - 1)^t}{(x + 3)^t} + 3 = {x^{{5^{m - 1}} - 3}} + {(x + 3)^{{5^{m - 1}} - 3}} + 3 = 0.$$ Multiplying both sides by ${x^3}{(x + 3)^3}$ yields
 \begin{align*}
{x^{{5^{m - 1}}}}{(x + 3)^3} + {x^3}{(x + 3)^{{5^{m - 1}}}} + 3{x^3}{(x + 3)^3} = 0.
\end{align*}
Raising both sides to the 5th power and simplifying, we obtain
$$x{(x + 3)^{15}} + {x^{15}}(x + 3) + 3{x^{15}}{(x + 3)^{15}}=0.$$
Furthermore, we factor this polynomial over $\mathbb{F}_5$ as:
\begin{equation*}
	\begin{split}
		3x(x + 3){(x + 4)^6}({x^4} + {x^3} + 3{x^2} + 2x + 1)({x^9} +4{x^7} + 4{x^6}\\
	+ 3{x^4} + 4x + 1)({x^9} + 2{x^8} + 3{x^7} + 3{x^6} + 2{x^2} + 4x + 3) = 0.
	\end{split}
\end{equation*}
Apparently, $x=1$ is a solution. For $x=2$, we have $\mu (x + 3) = \mu (5) = 0$, contradicting the requirement $\mu (x + 3) = 1$. Since $\gcd(4,m)=1$, the quartic factor remains irreducible over $\mathbb{F}_{{5^m}}$ by Lemma 2.4. From Lemma 2.3, the polynomials with degree 9 factor into products of three irreducible cubics over ${\mathbb{F}_{{5^3}}}$. Therefore, for odd $m$ with $9\nmid m$, the above equation has no solution in $\mathbb{F}_{5^m}^*\backslash \{ 1\} $ satisfying $\mu (x) = \mu (x + 3) = 1$.

(ii) Following a similar approach, we consider $${x^t} + {(x + 3)^t} + 3 =0.$$ This equation has solutions $x=1,2$. For $x = 1$, $\mu (x + 3) = \mu ( - 1) = 1$, contradicting $\mu (x + 3) =  - 1$. For $x = 2$, $\mu (x + 3) = \mu ( 0) = 0$, again a contradiction. Hence, there is no solution in $\mathbb{F}_{5^m}^*$ with $\mu (x) =1$ and $ \mu (x + 3) = -1$ when $m$ is odd and $9\nmid m$.

(iii) Now, we consider $${x^t} + {( - 1)^t}{(x - 3)^t} + 3 = {x^{{5^{m - 1}} - 3}} + {(x - 3)^{{5^{m - 1}} - 3}} + 3 = 0.$$ Simplifying as before leads to
$$x{(x - 3)^{15}} + {x^{15}}(x - 3) + 3{x^{15}}{(x - 3)^{15}} = 0.$$
Over $\mathbb{F}_5$, this factors as:
\begin{equation*}
	\begin{split}
	x{(x + 1)^6}(x + 2)({x^4} + 4{x^3} + 3{x^2} + 3x + 1)({x^9} + 4{x^7} + {x^6}\\
	 +2{x^4} + 4x + 4)({x^9} + 3{x^8} + 3{x^7} + 2{x^6} + 3{x^2} + 4x + 2) = 0.
	\end{split}
\end{equation*}
For $x=-1$ , $\mu (x) = \mu ( - 1) = 1$, contradicting $\mu (x) =  - 1$. For $x = 3$, $\mu (x - 3) = \mu (0) = 0$, contradicting $\mu (x - 3) =  - 1$. As in part (i), the irreducible factors of degrees $4$ and $9$ imply no additional solutions in $\mathbb{F}_{5^m}^*$ with $\mu (x) = \mu (x - 3) =- 1$ when $m$ is odd and $9\nmid m$. 

Therefore, all conditions of Theorem 4.2 are satisfied, and $\mathcal{C}_5(0,s,t)$ is optimal.\qed\\

\dse{4.3~~The third class of optimal quinary cyclic codes ${\mathcal{C}_5}(0,s,t)$ with minimum distance 4}

In this subsection, we discuss the optimality of the quinary cyclic code ${\mathcal{C}_5}(0,s,t)$, where $$s=\frac{5^m+1}{2} \text{ and } t=5^{\frac{m+1}{2}}-5^{\frac{m-1}{2}}+1,$$ and $m$ is an odd positive integer.\\

\noindent\textbf{Lemma 4.6.} Let $t=5^{\frac{m+1}{2}}-5^{\frac{m-1}{2}}+1$. If $m$ is odd, then $|C_t|=m$.\\

\noindent\textbf{Proof.} Note that for any $1\le l \le m-1$, $$\gcd(5^l-1,5^m-1)=5^{\gcd(l,m)}-1<5^{\frac{m-1}{2}}-1.$$ Therefore,
\begin{align*}
\gcd(t,5^m-1)\cdot \gcd(5^l-1,5^m-1) & \le t\cdot \gcd(5^l-1,5^m-1)\\
&<(5^{\frac{m+1}{2}}-5^{\frac{m-1}{2}}+1)\cdot(5^{\frac{m-1}{2}}-1)\\
&=4\cdot5^{m-1}-3\cdot5^{\frac{m-1}{2}}-1\\
&<5^m-1.
\end{align*}
By Lemma 2.2, it follows that $|C_t|=m$.\qed\\

\noindent\textbf{Theorem 4.7.} Let $t=5^{\frac{m+1}{2}}-5^{\frac{m-1}{2}}+1$ and $s=\frac{5^m+1}{2}$. Suppose that $m$ is an odd integer and $\gcd(5,m)=1$. Then $\mathcal{C}_5(0,s,t)$ has parameters $[5^m - 1, 5^m - 2m - 2, 4]$ and is optimal.\\

\noindent\textbf{Proof.} By Lemma 4.1, since $t\equiv1~(\bmod~4)$, it follows that $t\notin {C_s}$. Lemma 4.6 gives $|{C_t}| = m$, so $\dim({\mathcal{C}_5}(0,s,t))={5^m} - 2m - 2$. According to Theorem 4.2, it suffices to verify Conditions C1, C2 and C3.

(i) Raising the equation ${x^t} -{(x + 3)^t} + 3 = 0$ to the $5^{\frac{{m+1}}{{2}}}$-th power yields $$x^{5^{\frac{m+1}{2}}+4} -(x + 3)^{5^{\frac{m+1}{2}}+4} + 3 = 0.$$ 
A straightforward computation shows that
$$x^{5^{\frac{m+1}{2}}}(2{x^3}+4{x^2}+3x+1)+3(x^4+2{x^3}+4{x^2}+3x)=0,$$
which simplifies to
$$2x^{5^{\frac{m+1}{2}}}({x^3}+2{x^2}+4x+3)-2x(x^3+2{x^2}+4x+3)=0.$$
Factoring further, we obtain
$$2x^2(x+1)(x+2)(x+4)(x^{5^{\frac{m+1}{2}}-1}-1)=0.$$
Clearly, $x=1$ is a solution. If $x+1=0$, then $\mu(x+3)=\mu(2)=-1$ for odd $m$, contradicting  $\mu(x+3)=1$. If $x+2=0$, then $\mu(x)=\mu(-2)=-1$, contradicting $\mu(x)=1$. Since $\gcd(5^{\frac{m+1}{2}}-1,5^m-1)=4$, the roots of $x^{5^{\frac{m+1}{2}}-1}-1$ are exactly $x=\pm1,\pm 2$. The case $x=2$ leads to $\mu(x)=\mu(2)=-1$, again contradicting $\mu(x)=1$. Therefore, Condition C1 is satisfied.

(ii) Similarly, applying the $5^{\frac{m+1}{2}}$-th power to ${x^t} +{(x + 3)^t} + 3 = 0$ gives $$x^{5^{\frac{m+1}{2}+4}} +(x + 3)^{5^{\frac{m+1}{2}+4}} + 3 = 0.$$ This reduces to
$$x^{5^{\frac{m+1}{2}}}({x^4}+{x^3}+2{x^2}+4x+3)-(x^4+2{x^3}+4{x^2}+3x+2)=0.$$
Note that ${x^4}+{x^3}+2{x^2}+4x+3=({x^2}+2x+3)(x^2+4x+1)$, and both quadratic factors are irreducible over $\mathbb{F}_5$. For odd $m$, Lemma 2.4 ensures they remain irreducible over $\mathbb{F}_{5^m}$, so ${x^4}+{x^3}+2{x^2}+4x+3\ne0$ for all $x\in \mathbb{F}_{5^m}$. Thus, we may write
\begin{align}\label{eq:4.4}
x^{5^{\frac{m+1}{2}}}=\frac{{x^4}+2{x^3}+4{x^2}+3x + 2}{{x^4}+{x^3}+2{x^2} +4x + 3}.
\end{align}
Let ${f_1}(x) ={x^4} +2{x^3} + 4{x^2} + 3x + 2$ and ${g_1}(x) ={x^4} +{x^3} + 2{x^2} + 4x + 3$. Raising both sides of \eqref{eq:4.4} to the $5^{\frac{m+1}{2}}$-th power and using $(x^{5^{\frac{m+1}{2}}})^{5^{\frac{m+1}{2}}}={x^5}$, we obtain
\begin{align}\label{eq:4.5}
x^5=\frac{x^{4\cdot5^{\frac{m+1}{2}}+2x^{3\cdot5^{\frac{m+1}{2}}}+4x^{2\cdot5^{\frac{m+1}{2}}}+3x^{5^{\frac{m+1}{2}}} + 2}}{x^{4\cdot5^{\frac{m+1}{2}}}+x^{3\cdot5^{\frac{m+1}{2}}}+2x^{2\cdot5^{\frac{m+1}{2}}} +4x^{5^{\frac{m+1}{2}}} + 3}.
\end{align}
Substituting \eqref{eq:4.4} into \eqref{eq:4.5}, we have
$$f_1^4 + 2f_1^3{g_1} + 4f_2^4g_1^2 + 3f_1g_1^3+2g_1^4 - {x^5}(f_1^4 +f_1^3g_1+ 2{f_1^2}g_1^2 +4f_1g_1^3+3 g_1^4 )= 0.$$ 
Factoring this above equation over $\mathbb{F}_5$ using Magma gives irreducible factors including 
\begin{equation*}
	\begin{split}
		(x+3)(x^5+x^2+3x+2)(x^5+x^4+3x^3+2)~~~~~~\\
	(x^5+2x^4+3x^2+2x+1)(x^5+3x^4+3x^3+2x+4).
	\end{split}
\end{equation*}
If $x+3=0$, then $\mu(x+3)=0$, contradicting $\mu(x+3)=-1$. By Lemma 2.4 and the assumption $\gcd(5,m)=1$, the equation ${x^t} -{(x + 3)^t} + 3 = 0$ has no solution in $\mathbb{F}_{5^m}^*$ satisfying $\mu (x) =1$ and $ \mu (x + 3) =- 1$. Hence, Condition C2 holds.

(iii) For the equation ${x^t} -{(x - 3)^t} + 3 = 0$, raising to the $5^{\frac{m+1}{2}}$-th power gives $${x^{{5^{\frac{m+1}{2}}}+4}} -{(x - 3)^{{5^{\frac{m+1}{2}}+4}}} + 3 = 0.$$ Simplifying, we obtain
$$x^{5^{\frac{m+1}{2}}}({x^3}+3{x^2}+4x+2)-(x^4+3{x^3}+4{x^2}+2x+2)=0.$$
Note that ${x^3}+3{x^2}+4x+2=(x+1)(x+3)(x+4)$. If $x+1=0$, then $\mu(x)=\mu(-1)=1$, contradicting $\mu(x)=-1$. If $x+3=0$, then $\mu(x-3)=\mu(-1)=1$, contradicting $\mu(x-3)=-1$. If $x+4=0$, then $\mu(x)=\mu(1)=1$, again contradicting $\mu(x)=-1$. Thus,  ${x^3}+3{x^2}+4x+2\neq0$ under the given conditions, and we may write 
\begin{align}\label{eq:4.6}
x^{5^{\frac{m+1}{2}}}=\frac{x^4+3{x^3}+4{x^2}+2x+2}{{x^3}+3{x^2}+4x+2}.
\end{align}
Let ${f_2}(x) =x^4+3{x^3}+4{x^2}+2x+2$ and ${g_2}(x) ={x^3}+3{x^2}+4x+2$. Raising both sides of \eqref{eq:4.6} to the $5^{\frac{m+1}{2}}$-th power leads to a polynomial equation:
$$f_2^4 + 3f_2^3{g_2} + 4f_2^2g_2^2 + 2f_2g_2^3+2g_2^4 - {x^5}(f_2^3g_2 + 3{f_2^2}g_2^2 +4f_2g_2^3+2 g_2^4 )= 0.$$  Factoring over $\mathbb{F}_5$ via Magma yields irreducible factors including
\begin{equation*}
	\begin{split}
	(x^5+2x^4+x^3+4x^2+2x+4)(x^5+2x^4+4x^3+1)~~~~\\
	(x^5+3x^4+4x^2+4x+2)(x^5+3x^4+3x^3+x^2+4x+1).
	\end{split}
\end{equation*}
By Lemma 2.4 and $\gcd(5,m)=1$, the equation ${x^t} -{(x - 3)^t} + 3 = 0$ has no solution in $\mathbb{F}_{5^m}^*$. Therefore, Condition C3 holds.\qed\\
 
\dse{4.4~~The fourth class of optimal quinary cyclic codes ${\mathcal{C}_5}(0,s,t)$ with minimum distance 4}

This subsection investigates the optimality of the quinary cyclic code ${\mathcal{C}_5}(0,s,t)$, where $$s=\frac{5^m+1}{2} \text{ and } t = {5^{\frac{m + 1}{2}}} - 3,$$ with $m$ being an odd positive integer.\\

\noindent\textbf{Lemma 4.8.} Let $t = {5^{\frac{m + 1}{2}}} - 3$. If $m$ is odd, then $|{C_t}| = m$.\\

\noindent\textbf{Proof.} Note that for any odd $m$ with $1\le l \le m-1$, we have $$\gcd(5^l-1,5^m-1)=5^{\gcd(l,m)}-1<5^{\frac{m-1}{2}}-1.$$ It follows that
$$\gcd(t,5^m-1)\cdot \gcd(5^l-1,5^m-1)\leq t\cdot (5^{\frac{m-1}{2}}-1).$$
Substituting the expression for $t$, we obtain
\begin{align*}
t\cdot(5^{\frac{m-1}{2}}-1)=({5^{ \frac{m + 1}{2}}} - 3)\cdot(5^{\frac{m-1}{2}}-1)=5^m-1-4\cdot(2\cdot5^{\frac{m-1}{2}}-1).
\end{align*}
Since $4\cdot(2\cdot5^{\frac{m-1}{2}}-1)>0$, we conclude 
$$\gcd(t,5^m-1)\cdot \gcd(5^l-1,5^m-1)<5^m-1.$$
Therefore, by Lemma 2.2, it follows that $|C_t|=m$ by Lemma 2.2.\qed\\

\noindent\textbf{Theorem 4.9.} Let $m$ be an odd integer such that $\gcd (7,m) = 1$, and define 
$$t = 5^{\frac{m + 1}{2}} - 3 \text{ and } s = \frac{{5^m} + 1}{2}.$$ Then the quinary cyclic code $\mathcal{C}_5(0,s,t)$ has parameters $[5^m - 1, 5^m - 2m - 2, 4]$ and is optimal.\\

 \noindent\textbf{Proof.} By Lemmas 4.1 and 4.8, we have $t \notin {C_s}$ and $|{C_t}| = m$, so $\textrm{dim}({\mathcal{C}_5}(0,s,t))=5^m-2m-2$. Let $h=\frac{m + 1}{2}$. We now verify that the conditions of Theorem 4.2 are satisfied.
 
 (i) We first consider the equation $${x^t} + {(x + 3)^t} + 3 = {x^{{5^h} - 3}} + {(x + 3)^{{5^h} - 3}} + 3 = 0.$$ Clearly, $x = 1$ is a solution. For $x \ne 1$, multiplying both sides by ${x^3}{(x + 1)^3}$ yields 
 $${x^{{5^h}}}({x^3} + 2{x^2} + x + 1) = {x^3}({x^3} + 4{x^2} + 2x + 3).$$
 
Note that ${x^3} + 2{x^2} + x + 1 = (x + 4)({x^2} + 3x + 4)$, and ${x^2} + 3x + 4$ is irreducible over $\mathbb{F}_5$. Since $m$ is odd, Lemma 2.4 implies that ${x^2} + 3x + 4$ remains irreducible over ${\mathbb{F}_{{5^m}}}$, so ${x^3} + 2{x^2} + x + 1 \ne 0$ for all $x \in \mathbb{F}_{5^m}\backslash \{ 0,1\} $. Hence,
\begin{align}\label{eq:4.7}
x^{5^h} = \frac{x^5 + 2x^3 }{x^2 + 3x + 4}.
\end{align}
Since $2h = m + 1$, we have $x^{5^{2h}} = x^5$. Raising both sides of \eqref{eq:4.7} to the $5^h$-th power gives
\begin{align}\label{eq:4.8}
x^{5^{2h}}=\frac{x^{5\cdot5^{h}}+2x^{3\cdot5^h}}{x^{2\cdot5^h}+3x^{5^h}+4}.
\end{align}
Let ${f_1}(x) = {x^5} + 2{x^3}$ and ${g_1}(x) = {x^2} + 3x + 4$. Substituting \eqref{eq:4.7} into \eqref{eq:4.8} and using ${x^{{5^{2h}}}} = {x^5}$, we obtain
\begin{align}\label{eq:4.9}
x^{5} = \frac{f_1^5+2f_1^3g_1^2}{f_1^2g_1^3+3f_1g_1^4+4g_1^5}.
\end{align}
Clearing denominators yields
$$f_1^5+2f_1^3g_1^2-x^5f_1^2g_1^3-3x^5f_1g_1^4-4x^5g_1^5= 0.$$ 
Using Magma, we factor the left-hand side over $\mathbb{F}_5$ as
$$x^5(x+3)^5(x+4)(x^7 + 2x^6 + x^4 + 3x + 4)(x^7 + 4x^6 + 3x^5 + 4x^4 + 3x^3 + 4x^2 + 2x + 3)=0.$$ 
If $x+3=0$, then $x=2$, and $\mu(2)=-1$, contradicting $\mu(x)=1$. If $\gcd (7,m) = 1$, Lemma 2.4 implies that the irreducible factors of degree $7$ have no roots in $\mathbb{F}_5$. Therefore, ${x^t} + {(x + 3)^t} + 3 = 0$ has no solution in ${\mathbb{F}_{5^m}}\backslash \{ 0,1\} $.

(ii) If $x = 1$, then $\mu (x + 3) = \mu (1 + 3) = 1$, contradicting $\mu (x + 3) =-1$. A similar argument as in (i) shows that if $\gcd(7,m)=1$, then ${x^t} + {(x + 3)^t} + 3 = 0$ has no solution in $ \mathbb{F}_{{5^m}}^*$ with $\mu(x+3)=-1$.

(iii) Now, we consider $${x^t} + {(x - 3)^t} + 3 = {x^{{5^h} - 3}} + {(x - 3)^{{5^h} - 3}} + 3 = 0.$$ After simplification, we obtain
$${x^{{5^h}}}({x^3} + 3{x^2} + x + 4) = {x^3}({x^3} + {x^2} + 2x +2).$$
Note that ${x^3} + 3{x^2} + x + 4 = (x + 1)({x^2} + 2x + 4)$, and ${x^2} + 2x + 4$ is irreducible over $\mathbb{F}_5$. By Lemma 2.4, it remains irreducible over ${\mathbb{F}_{{5^m}}}$, so $x^2+ 2x + 4\neq0$ for all $x\in\mathbb{F}_{5^m}^*$. If $x+1=0$, then $\mu(x)=\mu( - 1) = 1$, contradicting $\mu(x) =  - 1$. For $x + 1 \ne 0$, we have
\begin{align}\label{eq:4.10}
{x^{{5^h}}} = \frac{x^3({x^3} + {x^2} + 2x + 2)}{{x^3} + 3{x^2} + x + 4} = \frac{x^3({x^2} + 2)}{{x^2} + 2x + 4}.
\end{align}
Raising both sides to the $5^h$-th power and using $x^{5^{2h}} = x^5$, we get 
\begin{align}\label{eq:4.11}
x^{5} = \frac{x^{3 \cdot {5^h}}({x^{2 \cdot {5^h}}} + 2)}{({x^{2 \cdot {5^h}}} + 2{x^{{5^h}}} + 4)}.
\end{align}
Let ${f_2}(x) = {x^2} + 2$ and ${g_2}(x) = {x^2} + 2x + 4$. Substituting \eqref{eq:4.10} into \eqref{eq:4.11} and clearing denominators leads to
$$f_2^5 + 2f_2^3g_2^2 - {x^5}f_2^2g_2^3 - 2{x^5}{f_2}g_2^4 - 4x^5g_2^5 = 0.$$ 
Factoring via Magma over $\mathbb{F}_5$ gives
$$x^5(x + 1){(x + 2)^5}({x^7} + {x^6} + 3{x^5} + {x^4} + 3{x^3} + {x^2} + 2x + 2)({x^7} + 3{x^6} + 4{x^4} + 3x + 1) = 0.$$  If $x + 1 = 0$, then $\mu (x) = 1$, a contradiction. If $x + 2 = 0$, then $x-3=0$, so $\mu (x - 3) = 0$, contradicting $\mu (x - 3) =  - 1$. By Lemma 2.4, the factors of degree $7$ have no roots in  $\mathbb{F}_{5^m}$ when $\gcd (7,m) = 1$. Hence, ${x^t} + {(x - 3)^t} + 3 = 0$ has no solution in  $\mathbb{F}_{5^m}^*$ with $\mu(x)=\mu(x-3)=-1$. 

In summary, under the condition that $m$ is odd and $\gcd (7,m) = 1$, the equations ${x^t} + {(x \pm 3)^t} + 3 = 0$ have no solutions in $\mathbb{F}_{5^m}^*\backslash \{ 1\} $. \qed\\

\noindent\textbf{Remark 5.} Table 3 displays the optimal quinary cyclic codes $\mathcal{C}_5(0,s,t)$ constructed in this paper. For comparison, Table 4 lists previously known codes $\mathcal{C}_5(0,s,t)$ with parameters $[5^m-1,5^m-2m-2,4]$, where $s=\frac{5^m+1}{2}$. Our codes are distinct from those in Table 4, confirming their novelty. Since $s=\frac{5^m+1}{2}$ is invertible modulo $5^m-1$, the cyclic code $\mathcal{C}_5(0,s,t)$ is equivalent to $\mathcal{C}_5(0,1,ts^{-1})$. Known quinary cyclic codes  $\mathcal{C}_5(0,1,t)$ with parameters $[5^m-1,5^m-2m-2,4]$ with the same parameters are provided in Table 5. A straightforward analytical verification shows that the codes constructed in this work are also inequivalent to those in Table 5, further demonstrating their new contribution.

\begin{table}[H]
\caption{Our main results $\mathcal{C}_5(0,\frac{5^m+1}{2},t)$ whose parameters are $[{5^m}-1,{5^m}-2m-2,4]$.}
\setlength{\tabcolsep}{0pt}
\small
\begin{tabular*}{14.4cm}{@{\extracolsep{\fill}}l l l l}
\hline
& $t$ & Conditions & Reference \\
\hline
1) & $7t\equiv -2\cdot{5^r}~(\bmod~{5^m}-1)$ & $m$ is odd and $0 \le r \le m-1$ & Theorem 4.3\\
2) & $t={5^{m-1}}-3$ & $m$ is odd and $9\nmid m$ & Theorem 4.5\\
3) & $t=5^{\frac{m+1}{2}}-5^{\frac{m-1}{2}}+1$ & $m$ is odd and $\gcd(5,m)=1$& Theorem 4.7\\
4) & $t=5^{\frac{m+1}{2}}-3$ & $m$ is odd and $\gcd(7,m)=1$ & Theorem 4.9\\
\bottomrule
\end{tabular*}
\end{table}

\begin{table}[H]
\caption{Known codes $\mathcal{C}_5(0,\frac{5^m+1}{2},t)$ whose parameters are $[{5^m}-1,{5^m}-2m-2,4]$.}
\setlength{\tabcolsep}{9pt}
\small
\begin{tabular}{@{\extracolsep{\fill}}l l l l}
\hline
& $t$ & Conditions & Reference \\
\hline
1) &$t={5^r}+2$ & $m=2r$, ${5^r}\equiv1~(\bmod~3)$ & [20, Theorem 2]\\
2) &$t=\frac{5^r+1}{2}$ & $m$ is even and $\gcd(r,2m)=1$ & [20, Theorem 2]\\
3) &$t={5^m}-2$ & $m$ is even  & [20, Theorem 2]\\
4) &$t=\frac{5^m-1}{2}+r$ & see Reference[20] & [20, Theorem 3, 4]\\
5) &$st\equiv2~(\bmod~{5^m}-1)$ & see Reference[20]  & [20, Theorem 5]\\
6) &$st\equiv3~(\bmod~{5^m}-1)$ & see Reference[20]  & [20, Theorem 6]\\
7) &$st\equiv7~(\bmod~{5^m}-1)$ & see Reference[20] & [20, Theorem 7]\\
\bottomrule
\end{tabular}
\end{table}

\begin{table}[H]
\caption{Known codes $\mathcal{C}_5(0,1,t)$ whose parameters are $[{5^m}-1,{5^m}-2m-2,4]$.}
\setlength{\tabcolsep}{2pt}
\small
\begin{tabular*}{14.4cm}{@{\extracolsep{\fill}}l l l l}
\hline
& $t$ & Conditions & Reference \\
\hline
1) &$t={5^r}+1$ & $0 \le r \le m-1$, $r\ne \frac{m}{2}$  & [6, Theorem 1]\\
2) &$x^t\in \mathbb{F}_{5^m}$ is PN or APN & $m\in \mathbb{Z}^+$ & [6, Theorem 6, 7]\\
3) &$t=\frac{5^m-1}{2}+3$ &  $\gcd(r,2m)=1$  & [6, Theorem 8]\\
4) &$t=\frac{5^m-1}{2}+\frac{5^r+1}{2}$ & $\gcd(r,2m)=1$ & [6, Theorem 8]\\
5) &$(5^m-2)t\equiv2\cdot5^r~(\bmod~5^m-1)$ & $m\ge 3$ is odd & [6, Theorem 9]\\
& & and $0 \le r \le m-1$ &\\
6) &$3t\equiv2\cdot5^r~(\bmod~5^m-1)$ & $m\ge 3$ is odd  & [6, Theorem 9]\\
& & and $0 \le r \le m-1$ &\\
7) &$t=\frac{5^m-1}{2}-1$ & $m\equiv0~(\bmod~2)$ & [6, Theorem 10]\\
8) &$t=\frac{5^m-1}{2}-3$ & $m\not\equiv0~(\bmod~2)$ & [6, Theorem 10]\\
9) &$t=14$ & $m\not\equiv0~(\bmod~2)$ & [6, Theorem 10]\\
10) &$t(5^h+1)\equiv5^r+1~(\bmod~5^m-1)$ & $\gcd(m,h+r)=1$ & [10, Theorem 15]\\
 & $t\equiv3~(\bmod~4)$ &and $\gcd(m,r-h)=1$ & \\
11) &$t(5^h-1)\equiv5^r-1~(\bmod~5^m-1)$ & $\gcd(m,h)=\gcd(m,r)=1$ & [10, Theorem 22]\\
 & $t\equiv2~(\bmod~4)$ or $t\equiv3~(\bmod~4)$ & and $\gcd(m,r-h)=1$ & \\
\bottomrule
\end{tabular*}
\end{table}

\dse{5~~Conclusions}
This work investigates the construction of optimal $p$-ary cyclic codes. First, by utilizing the properties of quadratic and quartic characters, we develop three new classes of optimal $p$-ary cyclic codes. Second, by analyzing the existence of solutions to specific equations over $\mathbb{F}_{5^m}^*$, we construct four new classes of optimal quinary cyclic codes.

\dse{Acknowledgments}
This research is supported by the National Natural Science Foundation of China (No.62201009) and the Anhui provincial Natural Science Foundation (No.21\\08085QA06).

\dse{Data availability}
No data was used for the research described in the article.


\begin{thebibliography}{99}
\bibitem{pa} R. Lidl, H. Niederreiter, Finite Fields, Encyclopedia of Mathematics and Its Applications, vol. 20, Cambridge University Press, 1997.

\bibitem{pa} C. Carlet, C. Ding, J. Yuan, Linear codes from highly nonlinear functions and their secret sharing schemes, IEEE Trans. Inf. Theory 51 (6) (2005) 2089-2102.

\bibitem{pa} C. Ding, T. Helleseth, Optimal ternary cyclic codes from monomials, IEEE Trans. Inf. Theory 59 (9) (2013) 2898–2904.

\bibitem{pa} X. Zeng, L. Hu, W. Jiang, Q. Yue, X. Cao, The weight distribution of a class of $p$-ary cyclic codes, Finite Fields Appl. 16 (1) (2010) 56–73.

\bibitem{pa} L. Wang, G. Wu, Several classes of optimal ternary cyclic codes with minimal distance four, Finite Fields Appl. 40 (2016) 126–137.

\bibitem{pa} G. Xu, X. Cao, S. Xu, Optimal $p$-ary cyclic codes with minimum distance four from monomials, Cryptogr. Commun. 8 (2016) 541–554.

\bibitem{pa} Y. Zhou, X. Kai, S. Zhu, J. Li, On the minimum distance of negacyclic codes with two zeros, Finite Fields Appl. 55 (2019) 134–150.

\bibitem{pa} D. Han, H. Yan, On an open problem about a class of optimal ternary cyclic codes, Finite Fields Appl. 59 (2019) 335–343.

\bibitem{pa} Y. Tian, Y. Zhang, Y. Hu, Optimal quinary cyclic codes with minimum distance four, J. Commun. 23 (8) (2019) 1293–1296.

\bibitem{pa} D. Liao, X. Kai, S. Zhu, P. Li, A class of optimal cyclic codes with two zeros, IEEE Commun. Lett. 23 (8) (2019) 1293-1296.

\bibitem{pa} Y. Liu, X. Cao, Four classes of optimal quinary cyclic codes, IEEE Commun. Lett. 24 (7) (2020) 1387–1390.

\bibitem{pa} J. Fan, Y. Zhang, Optimal quinary cyclic codes with minimum distance four, Chin. J. Electron. 29 (3) (2020) 515–524.

\bibitem{pa} H. Zhao, R. Luo, T. Sun, Two families of optimal ternary cyclic codes with minimal distance four, Finite Fields Appl. 79 (2022) 101995.
    
\bibitem{pa} Y. Liu, X. Cao, Optimal $p$-ary cyclic codes with two zeros, Appl. Algebr. Eng. Commun. Comput., 34 (2023) 129-138.
    
\bibitem{pa} Z. Ye, Q. Liao, On the Ding and Helleseth’s 7th open problem about optimal ternary cyclic codes, Finite Fields App. 92 (2023) 102284.

\bibitem{pa} G. Wu, H. Liu, Y. Zhang, Several classes of optimal $p$-ary cyclic codes with minimum distance four, Finite Fields Appl. 92 (2023) 102275.

\bibitem{pa} Y. Liu, X. Cao, Three new classes of optimal quinary cyclic codes with minimum distance four, Appl. Algebra Eng. Commun. Comput., https://doi.org/10.1007/s00200-023-00621-7.

\bibitem{pa12} T. Wu, L. Liu, L. Li, Several classes of optimal cyclic codes with three zeros, Appl. Algebra Eng. Commun. Comput., 36 (2025) 743-767.

\bibitem{pa} T. Wu, X. Zhu, L. Liu, Optimal quinary cyclic codes with three zeros, Cryptogr. Commun. 16 (2024) 801–823.

\bibitem{pa} J. Fan, X. Zeng, Optimal quinary cyclic codes with three zeros, Finite Fields Appl. 101 (2025) 102537.

\bibitem{pa} Z. Zha, L. Hu, G. Wu, New classes of optimal $p$-ary cyclic codes with minimum distance four, Finite Fields Appl. 103 (2025) 102588.
    
\bibitem{pa} Y. Liu, X. Cao, Z.  Zha, More classes of optimal quinary cyclic codes of form $\mathcal{C}_{(1,e,s)}$, Appl. Algebra Eng. Commun. Comput., 36 (2025) 327–339.

\end{thebibliography}
\end{document}